\documentclass[%
reprint,
noeprint,
superscriptaddress,
 amsmath,amssymb,
 aps,
floatfix
]{revtex4-2}
\usepackage[utf8]{inputenc}

\usepackage{graphicx}
\usepackage{float}

\usepackage                             { dsfont }
\usepackage                             {amsmath}
\usepackage                             {acronym}
\usepackage[colorlinks,linktocpage,bookmarks=false,citecolor=blue,linkcolor=red,urlcolor=blue]{hyperref}

\usepackage[normalem]{ulem}

\newacro{qft}[QFT]{quantum field theory}
\newacro{qed}[QED]{quantum electrodynamics}
\newacro{qcd}[QCD]{quantum chromodynamics}
\newacro{rg}[RG]{renormalization group}
\newacro{frg}[FRG]{functional renormalization group}
\newacro{de}[DE]{derivative expansion}
\newacro{ir}[IR]{infrared}
\newacro{uv}[UV]{ultraviolet}
\newacro{ssb}[SSB]{spontaneous symmetry breaking}
\newacro{qm}[QM]{quantum mechanics}
\newacro{lpa}[LPA]{local potential approximation}
\newacro{fft}[FFT]{fast Fourier transform}
\newacro{gsl}[GSL]{GNU Scientific Library}
\newacro{rk4}[RK4]{4th order Runge-Kutta}
\newacro{pms}[PMS]{principle of minimal sensitivity}
\newacro{cb}[CB]{conformal bootstrap}
\newacro{pmc}[PMC]{principal of maximal conformality}
\newacro{dmrg}[DMRG]{density matrix renormalization group}
\newacro{bmw}[BMW]{Blaizot, M\'endez-Galain and Wschebor}

\newcommand{\mi}{\mathrm i}
\DeclareMathOperator{\tr}{Tr}

\newcommand{\Gam}[1]{\Gamma^{(#1)}}

\begin{document}
    \title{Physical properties of the massive Schwinger model from the nonperturbative functional renormalization group}
    \author{Patrick Jentsch}
    \affiliation{Institut f\"ur Theoretische Physik Universit\"at Heidelberg,
    Philosophenweg 16, D-69120 Heidelberg}
\affiliation{Sorbonne Universit\'e, CNRS, Laboratoire de Physique Th\'eorique de la Mati\`ere Condens\'ee, LPTMC, F-75005 Paris, France}
    \author{Romain Daviet}
    \affiliation{Sorbonne Universit\'e, CNRS, Laboratoire de Physique Th\'eorique de la Mati\`ere Condens\'ee, LPTMC, F-75005 Paris, France}
    \author{Nicolas Dupuis}
    \affiliation{Sorbonne Universit\'e, CNRS, Laboratoire de Physique Th\'eorique de la Mati\`ere Condens\'ee, LPTMC, F-75005 Paris, France}
    \author{Stefan Floerchinger} 
    \affiliation{Institut f\"ur Theoretische Physik Universit\"at Heidelberg,
    	Philosophenweg 16, D-69120 Heidelberg}

\date{June 24, 2021} 
    
\begin{abstract}
    {
    We investigate the massive Schwinger model in $d=1+1$ dimensions using bosonization and the nonperturbative \acl{frg}. In agreement with previous studies we find that the phase transition, driven by a change of the ratio $m/e$ between the mass and the charge of the fermions, belongs to the two-dimensional Ising universality class. The temperature and vacuum angle dependence of various physical quantities (chiral density, electric field, entropy density) are also determined and agree with results obtained from \acl{dmrg} studies.  Screening of fractional charges and deconfinement occur only at infinite temperature. Our results exemplify the possibility to obtain virtually all physical properties of an interacting system from the \acl{frg}.}  
    
\end{abstract}
    
\maketitle

\section{Introduction}

Historically the renormalization-group approach has been used primarily for the study of universal properties of systems near a second-order phase transition~\cite{kadanoff1966,Wilson:1971bg,Wilson:1971dh,Wilson74}. The nonperturbative \acf{frg}, which is the modern implementation of Wilson's RG~\cite{Wetterich:1992yh,Ellwanger:1993mw,Morris:1993qb,WetterichTetradis,Delamotte12,Dupuis:2020fhh}, has been quite successful in this respect since it yields accurate values of the critical exponents associated with the Wilson-Fisher fixed point of O($N$) models~\cite{Balog19,DePolsi20}, comparable with the best estimates from field-theoretical perturbative RG~\cite{Guida98,Kompaniets17}, Monte Carlo simulations~\cite{Hasenbusch10,Campostrini06,Campostrini02,Hasenbusch19,Clisby16,Clisby17} or conformal bootstrap~\cite{Kos16,SimmonsDuffin17,Echeverri16,Chester20}.

However the FRG is not restricted to the study of universal properties. When the microscopic theory is known or with the help of additional input (e.g. the knowledge of some physical quantities of the microscopic theory), one can compute the expectation value of observables even when the system is far away from a critical point. The FRG has been used in many models of quantum and statistical field theory ranging from statistical physics and condensed matter to high-energy physics and quantum gravity~\cite{Dupuis:2020fhh}. Besides the interest in models where perturbative approaches or numerical methods are difficult for various reasons, there is an ongoing effort to characterize and quantify the efficiency of the FRG by considering well-known models of field theory.

In this paper, to exemplify the predictive power of the FRG formalism, we determine both universal and nonuniversal properties of the Schwinger model, at zero and finite temperature. Our analysis is based on a previous study of the sine-Gordon model \cite{DavietDupuis} and goes beyond previous \ac{frg} applications to the Schwinger model \cite{Nandori:2009ad,Nandori:2010ij}. The results compare well with numerical studies except for the finite-temperature phase diagram where, contrary to the expectation, we find a region with \ac{ssb} at $T> 0$. This observation is likely an artifact of the FRG calculation using a truncated derivative expansion, as we will discuss.

After briefly reviewing the Schwinger model in Sec. \ref{sec:schwinger}, we start our discussion in Sec. \ref{sec:mapping} by making the identification of the parameters between the fermionic and the bosonic theory precise. We then briefly review the \ac{frg} using a \ac{de} at finite temperature in Sec. \ref{sec:frg}. In Sec. \ref{sec:phase_transition} we investigate the phase transition of the Schwinger model and determine the critical ratio $(m/e)_c$ and the critical exponents. We discuss the observation of a nonconvex effective potential in the \ac{ssb}-phase in Sec. \ref{sec:convexity} and then, still at zero temperature, determine the $\theta$ dependence of nonuniversal observables in Sec. \ref{sec:ireval}. This calculation is extended to finite temperature in Sec. \ref{sec:finiteT} where we also consider the issue of \ac{ssb} at finite temperature. Finally we conclude the paper in Sec. \ref{sec:conclusion}.

\section{The Schwinger model}
\label{sec:schwinger}

The Schwinger model \cite{Schwinger:1962tp} describes \ac{qed} in $d=1+1$ space-time dimensions. It was originally introduced to show that a gauge field, for which an explicit mass term is forbidden by gauge invariance, can acquire a mass dynamically through the chiral anomaly. With both massless and massive fermions, it demonstrates confinement \cite{abdallah,Lowenstein:1971fc} and thus serves as an important toy model for real confining theories such as \ac{qcd} in 3+1 dimensions. 

The microscopic action of the Schwinger model in Euclidean space is given by~\cite{ZinnJustin:2002ru} 
%
\begin{align}
\label{eq:action_fermionic}
    S[\bar \psi, \psi, A]
    ={}& \int_x \biggl\{ -\bar \psi\gamma^\mu (\partial_\mu + \mi A_\mu)\psi - m\bar\psi\psi \nonumber \\ &  + \frac{1}{4e^2} F_{\mu\nu}F^{\mu\nu}+\mi\frac{\theta}{4\pi}  \epsilon^{\mu\nu}F_{\mu\nu} \biggr\}
\end{align}
%
where $A_\mu$ is the usual $U(1)$ gauge field and $F_{\mu\nu}=\partial_\mu A_\nu-\partial_\nu A_\mu$ the corresponding field strength. Moreover, $\psi$ and its conjugate $\bar\psi$ are two-component Dirac spinor Grassmann fields describing the charged fermions. The Dirac matrices and the real antisymmetric tensor $\epsilon_{\mu\nu}$ are taken in Euclidean space such that $\frac{1}{2}\{\gamma^\mu,\gamma^\nu\}=g^{\mu\nu} = \text{diag}(1,1)$ and $\epsilon_{10}=-\epsilon_{01}=\epsilon^{10}$. The two free parameters of this theory are the fermion mass $m$ and the electric charge $e$ (the latter has dimension of mass in $d=1+1$). Contrary to 3+1-dimensional \ac{qed}, the Schwinger model permits a topological term linear in the field strength, that gives rise to an additional free parameter, the vacuum angle $\theta$. As the model is a super-renormalizable theory (positive mass dimension of the coupling constant $e$), it can be defined without an explicit \ac{uv} cutoff ($\Lambda\rightarrow \infty$). 

At low energies, the Schwinger model can be mapped onto the massive sine-Gordon model~\cite{Coleman:1976uz} defined by the action
\begin{equation}
\label{eq:action_bosonic}
    S[\phi] = \int_x \left\{ \frac{1}{2}\partial^\mu\phi\partial_\mu\phi-u \cos(\beta \phi + \theta) + \frac{1}{2} M^2\phi^2\right\}.
\end{equation}
with $\beta=2\sqrt{\pi}$, the Schwinger mass $M=e/\sqrt \pi$  and the parameter $u$, linearly related to the fermion mass $m$. The bosonic action in Eq. (\ref{eq:action_bosonic}) is also written in Euclidean space. Contrary to the fermionic action, it is only welldefined with a \ac{uv} regularization and the relation between $m$ and $u$ depends on the implementation of this cutoff. The exact relation between these two parameters is not well documented for the path integral formalism in the literature, as the equivalence is most commonly proven in the operator formalism \cite{abdallah,Lowenstein:1971fc} or in mass perturbation theory \cite{Naon:1984zp} which both, at least implicitly, rely on a normal ordering choice \cite{Coleman:1974bu}. This freedom is of course absent in the path integral formalism and we shall discuss the exact relation in Sec. \ref{sec:mapping}. The equivalence between the theories in Eqns. (\ref{eq:action_fermionic}) and (\ref{eq:action_bosonic}) has also been proven at finite temperature \cite{Faruk:2018mcs}.

From the form of the bosonic action, Coleman first conjectured a continuous phase transition between a symmetric phase and a phase in which the charge conjugation symmetry $\phi\rightarrow-\phi$ at $\theta=\pi$ is spontaneously broken \cite{Coleman:1976uz}. In the \ac{ssb} phase ``half-asymptotic'' fermions can exist as widely separated kink-antikink pairs, as long as they are ordered in the right way. In the symmetric phase, no kink solutions exist. Fermions can only exist in bound meson states, so there is confinement in the weak sense. The byname weak refers to the fact that the fermion-antifermion pair in the bound state can still be separated at finite energy cost, through the creation of a new pair of particles from the vacuum. The newly created particles form new bound states with the original particles to screen their electric field. 
Fractionally charged external particles cannot be screened by pair creation~\cite{ColemanJackiw:1975pw, frishman_sonnenschein_2010} at nonzero fermion mass and therefore experience confinement in the strong sense, i.e. they can only be separated for infinite energy cost. In the massless model, arbitrary charges can be screened though.

Numerically the existence of a phase transition when $\theta=\pi$ was shown using various approaches~\cite{Hamer:1982mx, Schiller:1983sj,Byrnes:2002nv, Buyens:2017crb}. In recent years \acf{dmrg} techniques have become quite popular and provide the most accurate result for the critical value $(m/e)_c=0.3335(2)$~\cite{Byrnes:2002nv}. Furthermore, estimates for the critical exponents were calculated \cite{ Byrnes:2002nv,Shimizu:2014fsa}, showing that the phase transition belongs to the universality class of the two-dimensional Ising model. 

\begin{figure}
    \centering
    \includegraphics[width=0.8\linewidth]{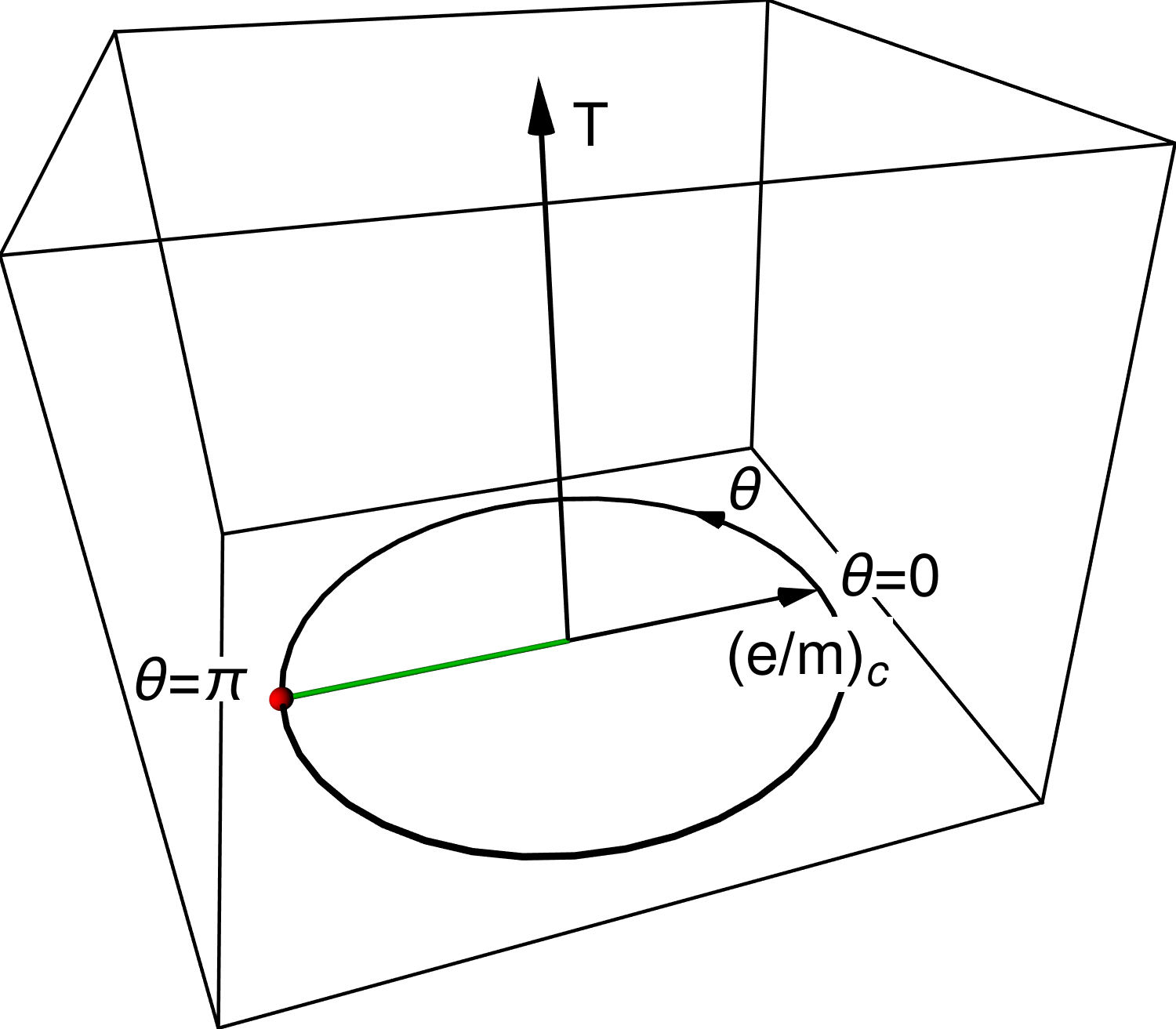}
    
    
    \caption{Schematic phase diagram of the Schwinger model depending on the vacuum angle $\theta$, ratio $e/m$ and temperature $T$. The green line marks the phase with \ac{ssb} and the red ball the critical point. Everywhere else the model has a unique ground state.}
    \label{fig:PhaseDiag3d}
\end{figure}

The \ac{dmrg} method also allowed the precise calculation of \ac{ir} observables, both at zero \cite{Buyens:2017crb} and finite temperature \cite{Banuls:2015ryj, Banuls:2016lkq, Buyens:2016ecr}. By the findings of Ref.~\cite{Buyens:2016ecr} we expect that at finite temperature there is no phase transition and the Schwinger model has a unique ground state, see Fig.~\ref{fig:PhaseDiag3d}.

\section{Mapping of Parameters}
\label{sec:mapping}

To deduce the parameters of the low-energy theory~(\ref{eq:action_bosonic}) from the microscopic model~(\ref{eq:action_fermionic}) one has to compare renormalized parameters or expectation values in the \ac{ir} obtained in the two approaches. The Schwinger model offers two limits which we know how to solve exactly and can be used to relate $u$, $M$ and $\Lambda$ to $m$ and $e$. One is the weak coupling limit $e/m\to 0^+$ or, in the bosonic formulation, $M\Lambda/u\to 0^+$ where the Schwinger model is equivalent to the sine-Gordon model at $K=\beta^2/8\pi=1/2$ (origin of Fig.~\ref{fig:PhaseDiag3d}). In the strong coupling limit $m/e\to 0^+$ or $u/M\Lambda\to 0^+$ (infinitely far into the radial direction of Fig. \ref{fig:PhaseDiag3d}), i.e. for massless fermions, the bosonic action is that of a free field. In the following two subsections, we show that both limits yield the same relation between the fermionic and bosonic parameters. 

\subsection{Weak coupling limit}

In the sine-Gordon model the soliton mass, i.e. the mass of the fermion, is known exactly \cite{Zamolodchikov:1995xk},
\begin{equation}
    m = b \Lambda \frac{2 \Gamma(\frac{K}{2-2K})}{\sqrt \pi \, \Gamma (\frac{1}{2-2K})}\left[ \frac{\Gamma(1-K)}{\Gamma(K)} \frac{\pi u}{2 (b\Lambda)^2} \right]^{1/(2-2K)} ,
\end{equation}
where $b$ is a nonuniversal scale factor that depends on the implementation of the UV cutoff~\cite{DavietDupuis}. 
For the special case $K=1/2$, which corresponds to the Schwinger model, this reduces to
\begin{equation}
    m = \frac{\pi u}{b \Lambda}.
\end{equation}
We assume a hard momentum cutoff at $p^2=\Lambda^2$, for which
\begin{equation}
    b = \frac{e^\gamma}{2},
\end{equation}
with the Euler-Mascheroni constant $\gamma\simeq0.5772$. A discussion of more general cutoff functions can be found in the supplemental material of Ref.~\cite{DavietDupuis}. The relation between $m$ and $u$ is then given by
\begin{equation}
\label{eq:relation}
    u = \frac{e^\gamma}{2\pi} \Lambda m.
\end{equation}

\subsection{Strong coupling limit}

Even though in the strong coupling limit there is neither a mass term in the fermionic model nor a cosine term in the bosonic model, we can still compare the relation between $u$ and $m$ through the expectation value of the chiral density. In the massless Schwinger model the expectation value for the chiral density at $\theta=0$ is given by \cite{Jayewardena:1988td,Sachs:1991en} 
\begin{equation}
\label{eq:chiraldensT0}
    \langle \bar\psi\psi\rangle = \frac{e^\gamma}{2\pi} \frac{e}{\sqrt\pi}.
\end{equation}
In the bosonic theory, this expectation value is obtained from the expectation value of the exponential operator
\begin{align}
    \langle \bar\psi\psi\rangle &= \frac{u}{m} \langle e^{\mathrm i \sqrt{4\pi}\phi}\rangle \nonumber \\
    &= \frac{u}{m} e^{-2\pi \Delta(0)} , 
\end{align}
where the average on the right-hand side is taken with the (Gaussian) action of the field $\phi$.  
The propagator $\Delta(x)=\langle\phi(x)\phi(0)\rangle$,
\begin{equation}
    \Delta(x) = \frac{1}{(2\pi)^2} \int \frac{\mathrm d^2p }{p^2+M^2}e^{\mathrm i px},
\end{equation}
must be calculated in the presence of the same cutoff that was used in the calculation in the weak coupling limit, i.e. a hard cutoff,
\begin{equation}
    \Delta(0) = \frac{1}{4\pi} \ln\left(\frac{M^2+\Lambda^2}{M^2}\right)= \frac{1}{2\pi} \ln\left(\frac{\Lambda}{M}\right) + O(\Lambda^{-2}).
\end{equation}
We deduce that
\begin{equation}
    u = \frac{\langle\bar\psi\psi\rangle}{\langle e^{\mathrm i \sqrt{4\pi}\phi}\rangle} m = \frac{e^\gamma}{2\pi} \Lambda m,
\end{equation}
which agrees with the result obtained from the weak coupling limit in Eq. (\ref{eq:relation}).

In the fermionic theory, there is no \ac{uv} cutoff, so it is evident that one of the parameters $e$ and $m$ sets the overall mass scale of the theory, while physics and especially the critical point of the phase transition may only depend on the ratio $m/e$. In the bosonic theory, this translates into the fact that, provided $\Lambda\gg e,m$, the expectation values and the critical point only depend on the ratio
\begin{equation}
\label{eq:map_ratios}
     \frac{m}{e} =\frac{2\sqrt\pi}{e^\gamma} \frac{u}{\Lambda M},
\end{equation}
and not, as one might naively expect, also on the dimensionless ratio $u/M^2$~\cite{note1}.

\section{Functional Renormalization Group Approach}
\label{sec:frg}

We apply the \ac{frg} approach to the bosonic form of the Schwinger model (\ref{eq:action_bosonic}) in Euclidean space where the imaginary-time dimension has a finite length given by the inverse temperature,
\begin{equation}
    \int_x = \int_{-\infty}^{\infty}\mathrm d r \int_{0}^{\frac{1}{T}}\mathrm d \tau ,
\end{equation}
with $x=(r,\tau)$. Wilson's idea of the \ac{rg} is realized by including a scale dependent \ac{ir} regulator $\Delta S_k[\phi]$ in the partition function
\begin{equation}
    \mathcal Z_k[J] = \int \mathcal D \phi e^{-S[\phi]-\Delta S_k[\phi] + \int_x J\phi}.
\end{equation}
The flowing effective action
\begin{equation}
    \Gamma_k[\phi] = -\ln \mathcal Z_k[J]+ \int_x J \phi - \Delta S_k[\phi],
\end{equation}
which is defined as the modified Legendre transform of $\ln \mathcal Z_k[J]$, then smoothly interpolates between the microscopic action $\Gamma_{k_\text{in}}[\phi] = S[\phi]$ for $k_\text{in}\gg \Lambda$ and the macroscopic quantum effective action $\Gamma_{k=0}[\phi]=\Gamma[\phi]$. To ensure this property $\Delta S_k[\phi]$ must behave as a masslike term of order $k$ for small momenta $p$ and Matsubara frequencies $\omega_n$, i.e. for $y=(p^2 + \omega_n^2/c_k^2)/k^2 \ll 1$, and become negligible for high energy modes $y \gg k^2$. Here $c_k$ is a flowing velocity that encodes the difference in scaling between momenta and frequency and will be defined below.

We realize this requirement by choosing 
\begin{equation}
    \Delta S_k[\phi] = \frac{1}{2}\int_q \phi(-q) R_k(q) \phi(q),
\end{equation}
with $q=(p,\omega_n = 2\pi T n)$ and the cutoff function
\begin{equation}
\label{eq:regulator}
    R_k(q) = Z_k k^2 y r(y) = Z_k k^2 y \frac{\alpha}{e^y-1} .
\end{equation}
We further define the combined momentum integral and Matsubara sum
\begin{equation}
    \int_q = T\sum_{n=-\infty}^{\infty} \int_{-\infty}^\infty \frac{\mathrm dp}{2\pi}.
\end{equation}
At $T=0$ the frequency sum turns into an integral. The prefactor $Z_k$ appearing in the cutoff function is defined below. With the free parameter $\alpha$ we can probe the dependence of the final results on the precise form of the regulator. 

The idea behind the \ac{frg} approach is to integrate Wetterich's exact flow equation \cite{Wetterich:1992yh,Ellwanger:1993mw,Morris:1993qb},
\begin{equation}
\label{eq:wetterich}
    \partial_k\Gamma_k[\phi]= \frac{1}{2} \tr\left\{ \partial_k R_k (\Gam 2_k[\phi] +R_k)^{-1} \right\},
\end{equation}
from the microscopic initial conditions $\Gamma_{k_\text{in}}$ to the effective action $\Gamma_{k=0}$. On the right hand side of equation (\ref{eq:wetterich}), $\Gam 2_k$ is the second-order functional derivative of $\Gamma_k$. 

This functional partial differential equation can usually not be solved exactly so we employ the \acl{de} which is a commonly used, systematic approximation scheme \cite{WetterichTetradis,Canet:2002gs,DePolsi20,Balog19}. At second order it relies on the following ansatz for the flowing effective action
\begin{equation}
    \Gamma_k[\phi]=\int_x \left\{ \frac{1}{2} Z_k(\phi)(\partial_x\phi)^2+\frac{1}{2} X_k(\phi)(\partial_\tau\phi)^2 + U_k(\phi) \right\} ,
\end{equation}
which is defined by three $k$-dependent functions of $\phi$: $Z_k(\phi)$, $X_k(\phi)$ and $U_k(\phi)$.
At finite temperature the $O(2)$ Euclidean symmetry is broken, such that we have to admit different renormalization functions for the spatial and temporal derivatives: $Z_k(\phi)\neq X_k(\phi)$. The initial conditions for these functions are fixed by the microscopic action (\ref{eq:action_bosonic}),
\begin{equation}
    Z_{k_\text{in}}(\phi) = 1, \hspace{1cm} X_{k_\text{in}}(\phi) = 1,
\end{equation}
\begin{equation}
    U_{k_\text{in}}(\phi)=\frac{1}{2} M^2 \phi^2 - u \cos(\sqrt{4\pi}\phi + \theta).
\end{equation}

One can split the effective potential into the mass term and a periodic contribution:  $U_k(\phi)=\frac{1}{2}M_k^{2}\phi^2 + V_k(\phi)$. Since Eq. (\ref{eq:wetterich}) only depends on the second derivative of $\Gamma_k$, which is periodic in the field, the periodicity of $Z_k(\phi)$, $X_k(\phi)$ and $V_k(\phi)$ is preserved during the \ac{rg} flow and the mass term does not renormalize:  $M_k = M$.

As in the study of the sine-Gordon model~\cite{DavietDupuis}, it is convenient to use the dimensionless functions 
\begin{equation}
\label{eq:def_couplingfcts}
\tilde Z_k(\phi) = \frac{Z_k(\phi)}{Z_k}, \quad \tilde X_k(\phi) = \frac{X_k(\phi)}{X_k} ,  \quad  \tilde U_k(\phi) = \frac{Z_k(\phi)}{Z_k k^2} ,
\end{equation}
where  
\begin{equation}
\label{eq:periodicnormalization}
Z_k = \langle Z_k(\phi) \rangle_\phi , \qquad 
X_k = \langle X_k(\phi) \rangle_\phi 
\end{equation}
and $\langle\cdots\rangle_\phi$ denotes the average over $[-\pi/\beta,\pi/\beta]$. We also define
\begin{equation}
\eta_k = -\partial_t \log Z_k, \hspace{1cm} \xi_k = -\partial_t \log X_k.
\end{equation}
One can also introduce dimensionless spacetime coordinates
\begin{equation}
\label{eq:rescalecoordinates}
\tilde r = k r \hspace{0.5cm} , \qquad \tilde \tau  = c_k k \tau ,
\end{equation}
where $c_k=\sqrt{X_k/Z_k}$ is a running velocity. The latter is however not the actual, physical, velocity 
\begin{equation}
c_{k,\text{phys}}^2=\frac{X_k(\phi_{0,k})}{Z_k(\phi_{0,k})}=c_k^2\frac{\tilde X_k(\phi_{0,k})}{\tilde Z_k(\phi_{0,k})} ,
\end{equation}
which is defined from the minimum of the potential: $U'_k(\phi_{0,k})=0$. In practice, the difference between $c_k$ and $c_{k,\rm phys}$ is small. 
Note that we do not rescale the field. This agrees with the fact that $\phi$ is an angular variable in the sine-Gordon and massive Schwinger models and has therefore a vanishing scaling dimension (as evident from the fact that $\phi$ appears inside a cosine).

These dimensionless functions however do not allow us to find a fixed point associated with a transition that belongs to the universality class of the two-dimensional Ising model. For example, it is well known that the field has scaling dimension $1/4$ at this transition, which seems to be incompatible with the angular character of $\phi$ in the massive Schwinger model. Thus, to obtain a fixed-point solution with nonvanishing scaling dimension of the field, one has to allow for a rescaling of the field, at least for fields near the origin, which is the important field range to understand a second-order transition. At the critical point (anticipating parts of the discussion in Sec.~\ref{sec:phase_transition}), we indeed find that the minimum of the potential vanishes as $\phi_{0,k}\sim 1/\sqrt{Z_k(0)}\sim k^{\eta/2}$ with $\eta=\lim_{k\to 0}\eta_k$, so that the rescaled variable $\tilde\phi_{0,k}=\sqrt{Z_k(0)}\phi_{0,k}$ reaches a fixed-point value $\tilde\phi_0^*$. This clearly shows that $Z_k(\phi_{0,k})$ has the usual meaning of a field renormalization factor near the transition and $\eta$ can be identified with the anomalous dimension. In practice, we define the rescaled field $\tilde\phi$ by  
	\begin{equation}
	\label{eq:phitilde} 
	\phi(\tilde \phi) = \tilde \phi + \left(\sqrt{\frac{Z_{k_{\text{in}}}}{Z_k}}-1 \right) \frac{\sin (\beta \tilde \phi)}{\beta} .
	\end{equation}
and redefine $Z_k$ and $X_k$ as
\begin{equation}
\label{eq:pointnormalization}
Z_k = Z_k(0), \qquad X_k = X_k(0).
\end{equation}
The change of variable~(\ref{eq:phitilde}) preserves the periodicity of the flow equations whereas $\phi\sim\tilde\phi/\sqrt{Z_k}$ for $\phi\to 0$. The redefinition~(\ref{eq:pointnormalization}) is necessary to identify $\eta=\lim_{k\to 0}\eta_k$ with the anomalous dimension. From a practical point of view, the rescaled field  allows us to zoom in on the small-field region even using an equally spaced grid in $\tilde\phi$, which is necessary to study the vicinity of the critical point where $\phi_{0,k}$ becomes extremely small. The change in the definition of $Z_k$ effectively only changes the regulator's amplitude, i.e. $\alpha$, to which the \ac{rg} flow is insensitive far from the critical point. Near the critical point $\alpha$ needs to be optimized anyway (see Sec.~\ref{sec:phase_transition} and Appendix~\ref{sec:pms}) and the change in the definition of $Z_k$ only affects the optimal value of $\alpha$. We emphasize that the nonlinear transformation~(\ref{eq:phitilde}) is only employed when looking at the phase transition. All expectation values in Secs. \ref{sec:ireval} and \ref{sec:finiteT} are obtained using a regular grid in $\phi$.

The flow equations for the functions $U_k$, $Z_k$ and $X_k$ can be obtained from Eq.~(\ref{eq:wetterich}). They differ from those of the sine-Gordon model~\cite{DavietDupuis} only in the initial conditions. For the determination of the function $X_k$ a discrete derivative, using the smallest Matsubara frequency, is used. The numerical results confirm that this choice performs better than a continuous derivative. We display concrete expressions for the flow equations in Appendix~\ref{sec:flow_equations}.

\section{Phase Transition}
\label{sec:phase_transition}

We study the continuous phase transition at zero temperature, i.e. on the $T=0$ plane of Fig.~\ref{fig:PhaseDiag3d}. By choosing $\theta=\pi$, we further restrict ourselves to the axis of the phase diagram where only the parameter $u/M\Lambda$ needs to be fine-tuned to arrive at the critical point (red dot of Fig.~\ref{fig:PhaseDiag3d}). This point is defined by an infinite correlation length, i.e., $\lim_{k\rightarrow 0} U_k^{\prime\prime}(0)/Z_k(0)=0$. Using Eq. (\ref{eq:map_ratios}) we find the critical ratio
\begin{equation}
    \left(\frac{e}{m}\right)_c = 0.318(13),
\end{equation}
which is in good agreement with the most accurate result found in the literature $(e/m)_c = 0.3335(2)$ \cite{Byrnes:2002nv}. 

We also determine the critical exponents $\eta$ and $\nu$. The anomalous dimension $\eta$ has been defined in the previous section. The exponent $\nu$ characterizes the divergence of the correlation length and can be obtained from the flow of $\tilde U_k(\tilde\phi)$ near the critical point and for $\tilde\phi\sim\tilde\phi_0$: 
\begin{equation}
    \partial_t \tilde U_k(\tilde\phi) \simeq A(\tilde\phi) e^{-\frac{t}{\nu}}
\end{equation}
at long RG time $|t|$ ($t=\ln(k/\Lambda)$). 
The exponents were optimized using the \ac{pms} on the parameter $\alpha$. More details can be found in Appendix \ref{sec:pms}.

Our results agree with the theoretical values of the Ising model in two dimensions, see Table \ref{tab:crit_exponents}, to an accuracy that is typical of a second-order \ac{de} calculation within the FRG \cite{DePolsi20,Dupuis:2020fhh}. We conclude, in accordance with previous studies~\cite{Byrnes:2002nv,Shimizu:2014fsa}, that the critical point of the Schwinger model belongs to the two-dimensional Ising universality class.

\begin{table}
    \centering
    \begin{tabular}{ccc}
    \hline\hline
         & $\eta$ & $\nu$  \\
    \hline
    Schwinger model (FRG-DE2)     &  $0.289(4)$ & $0.949(12)$ \\
    2D Ising (FRG-DE2) & $0.289(4)$ & $0.946(11)$  \\ 
    2D Ising (exact) \cite{Pelissetto:2000ek} & $1/4$ & $1$ \\
    \hline
    \end{tabular}
\caption{Critical exponents $\eta$ and $\nu$ in the massive Schwinger model obtained from the FRG approach and compared to those of the two-dimensional Ising model. The error due to the DE2 truncation is not included. }
\label{tab:crit_exponents}
	
\end{table}

\section{Convexity in the Ordered Phase}
\label{sec:convexity}

\begin{figure}
    \centering
    \includegraphics[width=\linewidth]{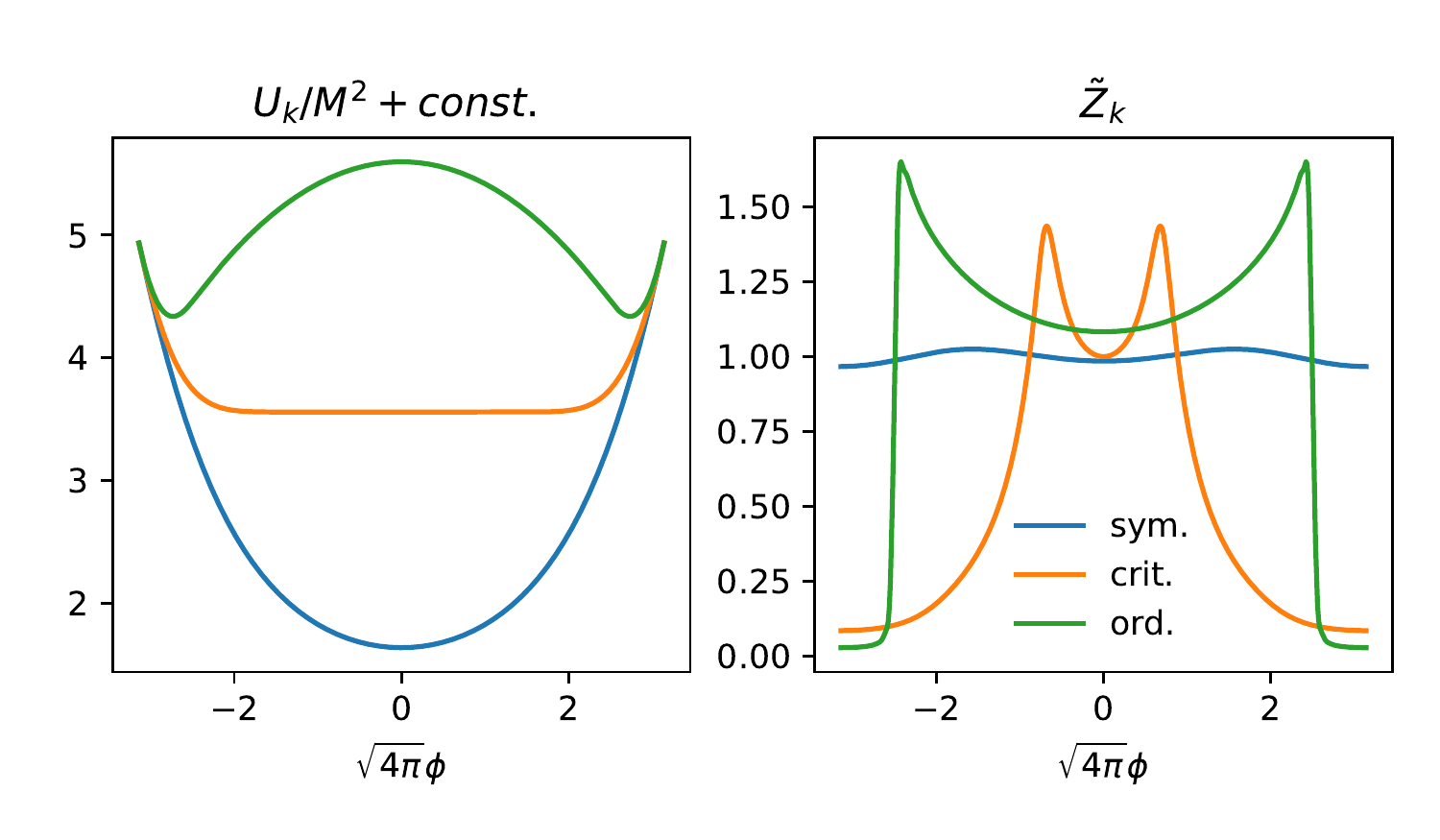}
    \caption{Effective potential $U_k(\phi)$ (up to a constant) and the function $\tilde Z_k(\phi)$ for $k\rightarrow0$ in the symmetric and ordered phase as well as near the critical point. Note that in the ordered phase the potential has not become convex and $\tilde Z_k$ develops a nonanalytic structure. $\tilde Z_k$ is normalized according to Eq.~(\ref{eq:pointnormalization}) near criticality and according to Eq.~(\ref{eq:periodicnormalization}) otherwise.}
    \label{fig:potentials}
\end{figure}

In the following, we want to determine nonuniversal expectation values in the \ac{ir}, where the effective action should have converged to a convex functional. However as can be seen in Fig.~\ref{fig:potentials}, the effective potential does not become convex in the ordered phase. This is caused by $\eta_k$ tending towards $2$ (from below) on the concave part of the potential. As a consequence $Z_k(\phi)$ is unbounded in this region, while on the convex part $\eta_k$ reaches zero and thus $Z_k(\phi)$ a finite value, leading to nonanalyticity in $\tilde Z_k$ near the potential minima. 

More importantly, unless $\eta_k$ converges to $2$ extremely slowly, i.e. $\lim_{k\to 0}(2-\eta_k)\ln k\to -\infty$, which is not realized in practice, the regulator function $R_k$ will approach a finite value in the \ac{ir}, $R_k =Z_k k^2 y r(y) \sim k^0 y r(y)$ for fixed $y$ and $k\rightarrow 0$. This violates our initial requirement, $R_{k=0}(q)=0$, that ensured the convergence of $\Gamma_k$ to the quantum effective action of the Schwinger model, and allows the effective potential to remain concave. On the other hand, it is not possible to choose a point of renormalization $Z_k=Z_k(\phi_r)$ with $\phi_r$ on the convex part of the potential. Since $Z_k(\phi)$ is unbounded, the regulator cannot ensure the positivity of the propagator and a pole is hit at finite \ac{rg} time. Note that, except in a few cases~\cite{Tetradis92,Tetradis96,Pelaez:2015nsa}, the DE does not in general guarantee the convexity of the effective action $\Gamma_{k=0}$ and nonconvex potentials have been obtained in other models~\cite{[{See, e.g., }]Debelhoir16b}.
So far it is unclear whether a regulator can be chosen such that both these problems can be avoided.

However, the concavity is only conceptually problematic. In practice, the mass $M$ effectively stops the flow of physical observables at a finite scale where the lack of convexity and the pathological small-$k$ behavior of $R_k$ are irrelevant. Many \ac{ir} observables can therefore be accurately predicted as shown below.

\section{Dependence of Observables on Vacuum Angle $\boldmath\theta$}
\label{sec:ireval}

In this section, we present our results for various observables at $T=0$ as a function of the vacuum angle $\theta$. Our discussion thus extends to the entire bottom plane of Fig.~\ref{fig:PhaseDiag3d}. We compare with the \ac{dmrg} results of Ref.~\cite{Buyens:2017crb}.

The string tension $\sigma_Q$, which characterizes confinement, is defined as the change of energy per unit length when two probe charges $\pm Q$ are introduced in the system. In the limit where the two charges are infinitely separated, they generate an additional background electric field that can be taken into account by a redefinition of the vacuum angle: $\theta\to\theta+\delta$ where $\delta=2\pi Q/e$. The string tension can thus be obtained by comparing two ground-state energies at different $\theta$. Since the energy per unit length is nothing but the effective potential in the FRG formalism, one obtains 
%
%
\begin{equation}
\sigma_Q = U_{\theta+\delta}(\phi_{0,\theta+\delta}) - U_{\theta}(\phi_{0,\theta}) , 
\end{equation}
where the notation emphasizes that the $k=0$ effective potential $U_\theta$ depends on the vacuum angle. Here, $\phi_{0,\theta}$ is the value of the field at the minimum of $U_\theta$.
The string tension is a periodic function of $\theta$. In the following we restrict ourselves to $\theta=0$ and consider (with the relabeling $\delta\to\theta$) 
%
\begin{equation}
\sigma_\theta = U_{\theta}(\phi_{0,\theta}) - U_{0}(\phi_{0,0}) .
\end{equation}

As the order parameter, the electric field $E = -\frac{e}{2\pi} \phi_{0,\theta}$ is also an interesting observable.

The chiral density, which serves as an order parameter for chiral symmetry breaking, is of particular interest in the massless Schwinger model since it is nonvanishing due to the chiral anomaly. In the massive model, the chiral symmetry is broken explicitly. Nevertheless, the anomaly still leaves an imprint on the chiral condensate. It can be obtained by taking a derivative of the partition function with respect to $u$ or, equivalently, the fermion mass. In terms of the effective potential this yields
\begin{equation}
\label{eq:chiraldensity} 
    \langle\bar\psi\psi\rangle=-\frac{\partial U_{\theta}(\phi_{0,\theta})}{\partial m} = \frac{e^\gamma}{2\pi}\Lambda\langle\cos(\sqrt{4\pi}\phi+\theta)\rangle.
\end{equation}
This derivative can be obtained either by taking the finite difference after integrating the flow equations or, as we discuss in Appendix~\ref{sec:flowchiral}, by integrating the flow equation for $\partial_m U_k$ but the results are less accurate.

Finally, we are interested in the particle spectrum. This was discussed originally by Coleman \cite{Coleman:1976uz}, who counted the number of stable particles in all phases. Unfortunately only the first excitation is available in the \acl{de} approximation. Its mass is given by the pole of the propagator, 
\begin{equation}
    M_1^2 = \lim_{k\rightarrow 0} \left.\frac{U_{\theta,k}^{\prime\prime}(\phi)}{Z_{\theta,k}(\phi)}\right|_{\phi=\phi_{0,\theta}}.
\end{equation}

\begin{figure}
    \centering
    \includegraphics[width=\linewidth]{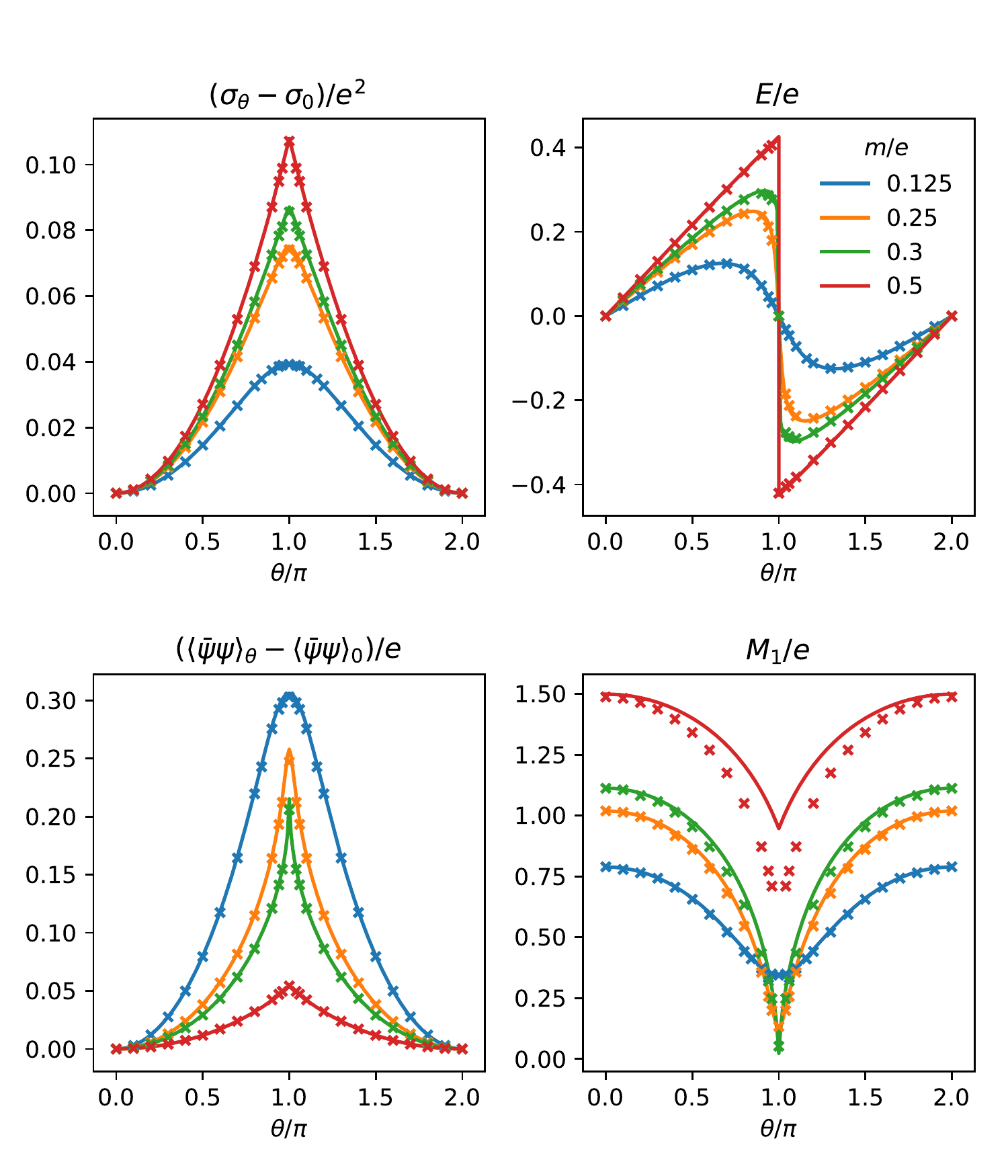}
    \caption{The infrared expectation values of the string tension, electric field, chiral density and the mass of the first excitation as a function of the vacuum angle $\theta$ (lines). The results are compared with the \ac{dmrg} results of Buyens \textit{et al.}~\cite{Buyens:2017crb} (crosses).}
    \label{fig:irexpvals}
\end{figure}

These four observables are shown as a function of $\theta$ in Fig.~\ref{fig:irexpvals} where they are compared with the DMRG results of Buyens \textit{et al.}~\cite{Buyens:2017crb}. The string tension and the electric field show an excellent agreement. From this, we can conclude that the effective potential has very well converged on its convex part, even though part of it remains concave. 

The mass gap is also very well reproduced in the symmetric phase. In the broken phase, there is a significant difference, especially for $\theta$ near its critical value $\theta_c=\pi$. Since the convex part of the effective potential seems to have converged to its infrared value, it seems that the problem must come from $Z_k(\phi)$. Indeed we see in Fig. \ref{fig:potentials} that in the \ac{ssb} phase $ Z_k(\phi)$ approaches a nonanalytic function. While it remains regular for any finite $k$, its exact structure exhibits pronounced peaks, which are hard to resolve numerically. This is less of an issue in the sine-Gordon model; the minimum being always at $\phi=0$, it is maximally far away from the difficult region. The error on the mass, which is about 10\% in the sine-Gordon model for $K=\frac{1}{2}$ \cite{DavietDupuis}, gets amplified to more than 50\% at $\theta=\pi$ by the mass term of the Schwinger model, which pushes the minima of $U_k(\phi)$ towards the peaks of $Z_k(\phi)$.

Numeric results for the string tension, electric field, and chiral density using matrix product states were also reported in Ref.~\cite{funcke}. They are compatible with ours but concern only the symmetric phase.

Our results are robust against a change of the regulator amplitude $\alpha$. Note that $\alpha$ must be chosen greater than $2$ to ensure that the propagator remains positive definite~\cite{Pelaez:2015nsa}.

As we expect, the electric field, as the order parameter of the phase transition, is discontinuous in the ordered phase for $\theta = \pi$. The string tension exhibits a cusp at this point. A larger fermion mass allows for a larger electric field and also stronger confinement. The larger string tension is easy to interpret as heavier fermions imply a higher threshold for pair production, responsible for a partial screening of fractional charges. The same is true for the electric field, as the quark-antiquark pairs try to screen the background electric field. Both quantities lose their $\theta$ dependency in the limit $m\rightarrow 0$ as is expected from the massless Schwinger model. This also holds for the mass gap $M_1$, which approaches the Schwinger mass $M$ in this limit. Near the phase transition, the theory becomes almost massless at $\theta=\pi$. For the chiral density, the opposite is true. In the zero-mass limit, it is strongly dependent on the vacuum angle due to the chiral anomaly. At finite mass, the chiral symmetry is explicitly broken, but in the weak coupling limit $m/e\gg 1$, the theory becomes a free fermionic theory, where the gauge field and therefore also the vacuum angle become unimportant.

\section{Temperature Dependence of Observables}
\label{sec:finiteT}

Finally we extend the discussion to finite temperatures, i.e. the vertical dimension of Fig.~\ref{fig:PhaseDiag3d}.

\begin{figure}
    \centering
    \includegraphics[width=\linewidth]{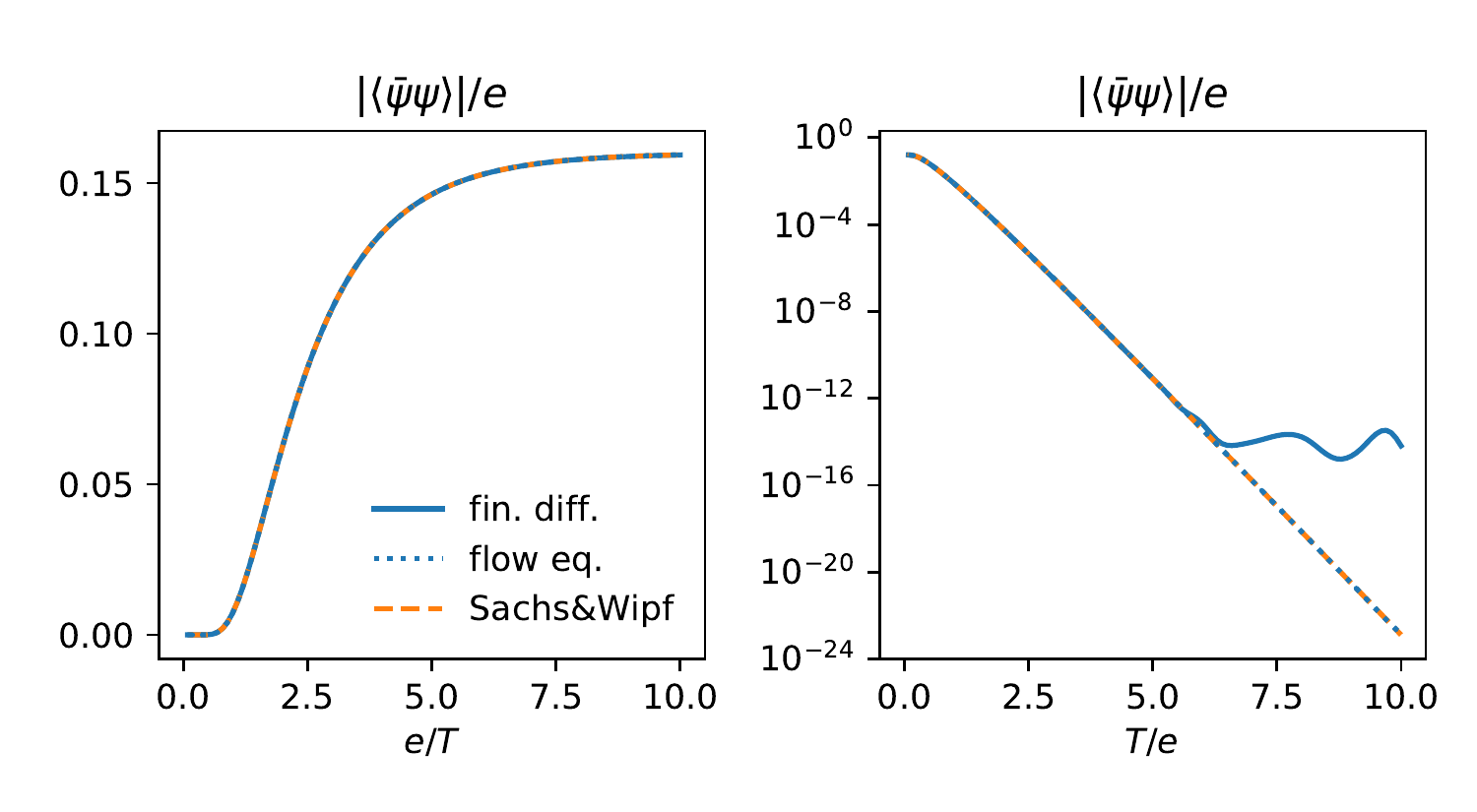}
    \caption{Result of our numerical calculation for the chiral condensate at finite temperature and $m=0$, using both the flow equation and the finite difference method, compared with the analytic result~(\ref{eq:sachswipf}) of Sachs and Wipf \cite{Sachs:1991en}.}
    \label{fig:sachswipf}
\end{figure}

In Fig.~\ref{fig:sachswipf} we benchmark our finite temperature implementation by comparing our result for the chiral density at $m=0$ with the exact analytic expression by Sachs and Wipf \cite{Sachs:1991en},
\begin{equation}
\label{eq:sachswipf}
    \langle\bar\psi\psi\rangle = \frac{M}{2\pi} e^\gamma e^{2I\left(\frac{M}{T}\right)} , \quad  I(x)=\int_0^\infty \frac{\mathrm d t}{1-e^{x \cosh t}}.
\end{equation}
Our result agrees both in the zero-temperature limit, where it approaches the value mentioned in Eq. (\ref{eq:chiraldensT0}), and in the high-temperature limit, where the chiral density decays exponentially,
\begin{equation}
    \langle\bar\psi\psi\rangle \xrightarrow{T\rightarrow \infty} 2 T e^{-\frac{\pi T}{M}}.
\end{equation}
At a value of about $T/e\sim 6$ machine precision is reached, so the finite difference method used to compute the derivative in~(\ref{eq:chiraldensity}) breaks down at this point. When using the calculation based on the flow equation (see Appendix~\ref{sec:flowchiral}), the lack of numerical precision manifests itself as a constant offset of the function $\partial_m U_{\theta}(\phi)$ which can be easily removed and the calculation therefore continued beyond this limit.

\begin{figure}
    \centering
    \includegraphics[width=\linewidth]{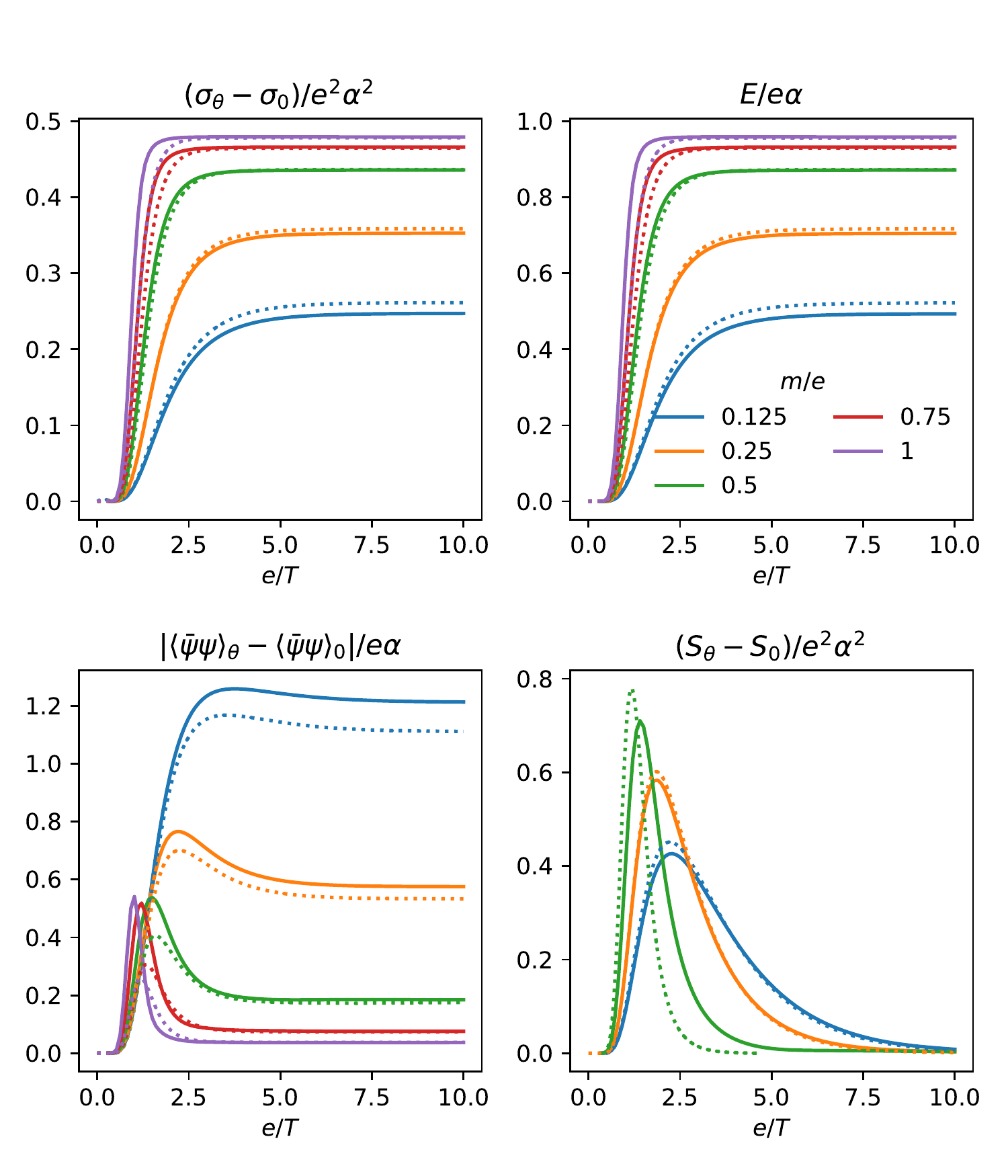}
    \caption{Dependence of string tension, electric field, chiral density and entropy density on inverse temperature for $\alpha=\theta/2\pi=0.05$ (continuous) compared with the \ac{dmrg} results of Buyens \textit{et al.} \cite{Buyens:2016ecr} (dotted lines). Both the chiral and entropy density are calculated using a finite difference with respect to mass and temperature respectively. }
    \label{fig:irtemp}
\end{figure}

\begin{figure}
    \centering
    \includegraphics[width=\linewidth]{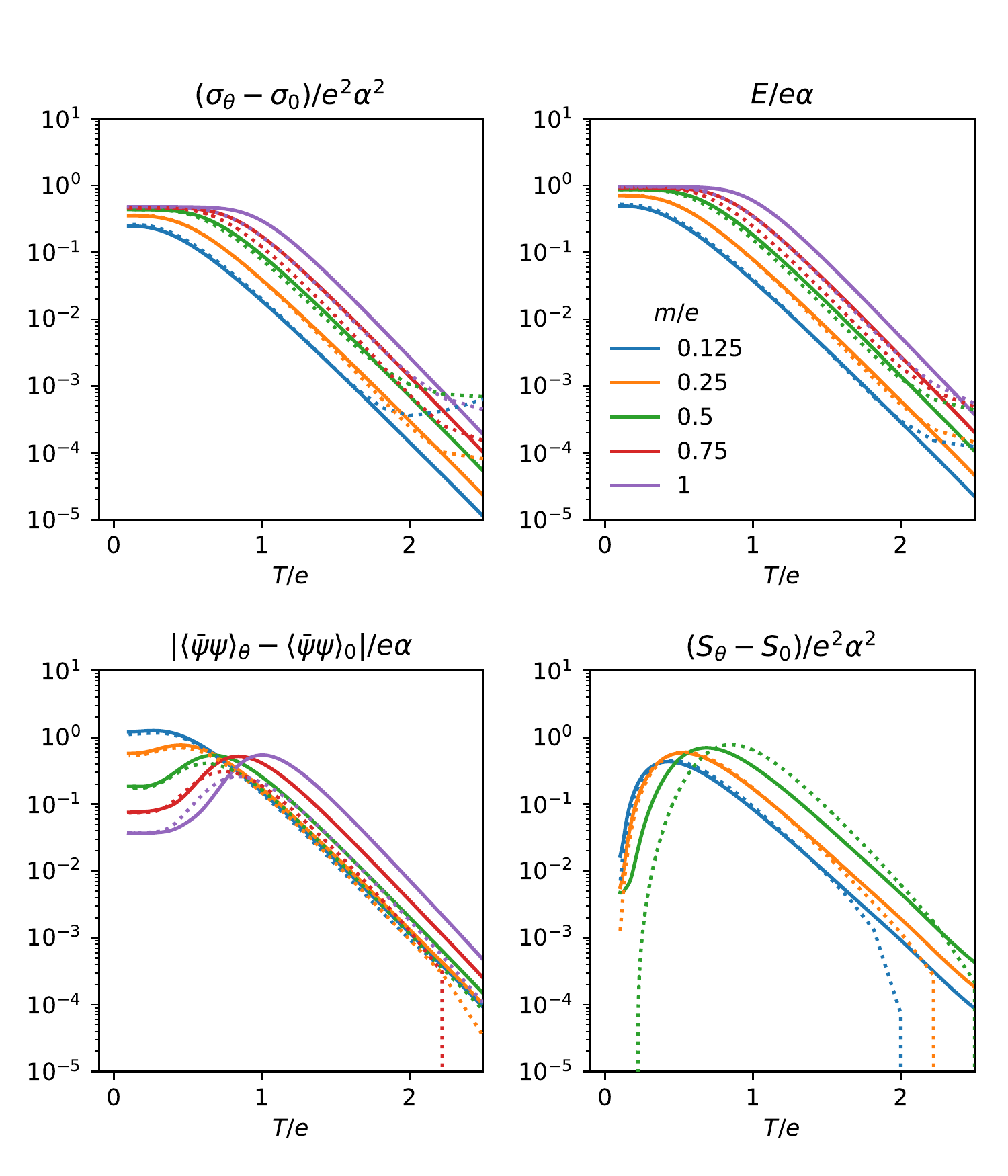}
    \caption{The same quantities as in Fig.~\ref{fig:irtemp} are shown but plotted logarithmically over temperature, such that the exponential decay with $T$ becomes apparent.
    For clarity the results are only plotted until $T/e\sim 2.5$, where Buyens' results (dotted lines) start to lose accuracy \cite{Buyens:2016ecr}.}
    \label{fig:irtemplog}
\end{figure}

We now consider nonzero values of $m$ where the string tension and the electric field are nonvanishing. Additionally, we will also calculate the entropy density, which is related to the string tension by
\begin{equation}
    S_\theta - S_0 = -\frac{\partial}{\partial T} \sigma_\theta ,
\end{equation}
using the finite difference method with $\Delta T=10^{-7} \Lambda$ to compute the temperature derivative. We first consider the limit $\theta\rightarrow 0$. To allow for nontrivial electric field, we consider a small, but nonvanishing, value $\alpha = \theta/2\pi = 0.05$. Our results in this limit are shown in Figs.~\ref{fig:irtemp} and \ref{fig:irtemplog}. They compare rather well with the finite-temperature \ac{dmrg} study by Buyens \textit{et al.}~\cite{Buyens:2016ecr}. In the low-temperature regime, the agreement is better for higher values of $m/e$. However, we believe that the FRG results are more reliable than the DMRG ones for all values of $m/e$ in this limit; our small, finite temperature results converge to our zero-temperature result, which are in accurate agreement also with Buyens' more recent zero-temperature study~\cite{Buyens:2017crb}, as we have shown in Fig.~\ref{fig:irexpvals}. 

In the high-temperature limit, our results agree very well with Ref.~\cite{Buyens:2016ecr} for the smaller values of $m/e$, but for higher values the amplitude is no longer accurately reproduced. Nevertheless the correct exponential decay is exhibited for all values of $m/e$ (Fig.~\ref{fig:irtemplog}). As compared to Ref.~\cite{Buyens:2016ecr} we can continue our results to higher temperatures, but eventually fail at the threshold of numerical precision as discussed before for the case $m=0$. The difference with the results of Ref.~\cite{Buyens:2016ecr} is more pronounced in the crossover region between the low- and high-temperature regimes, especially for the chiral condensate. For the three higher values of $m/e$, our results for the peaks are systematically higher, even though the zero temperature limits agree. 

\begin{figure}
    \centering
    \includegraphics[width=\linewidth]{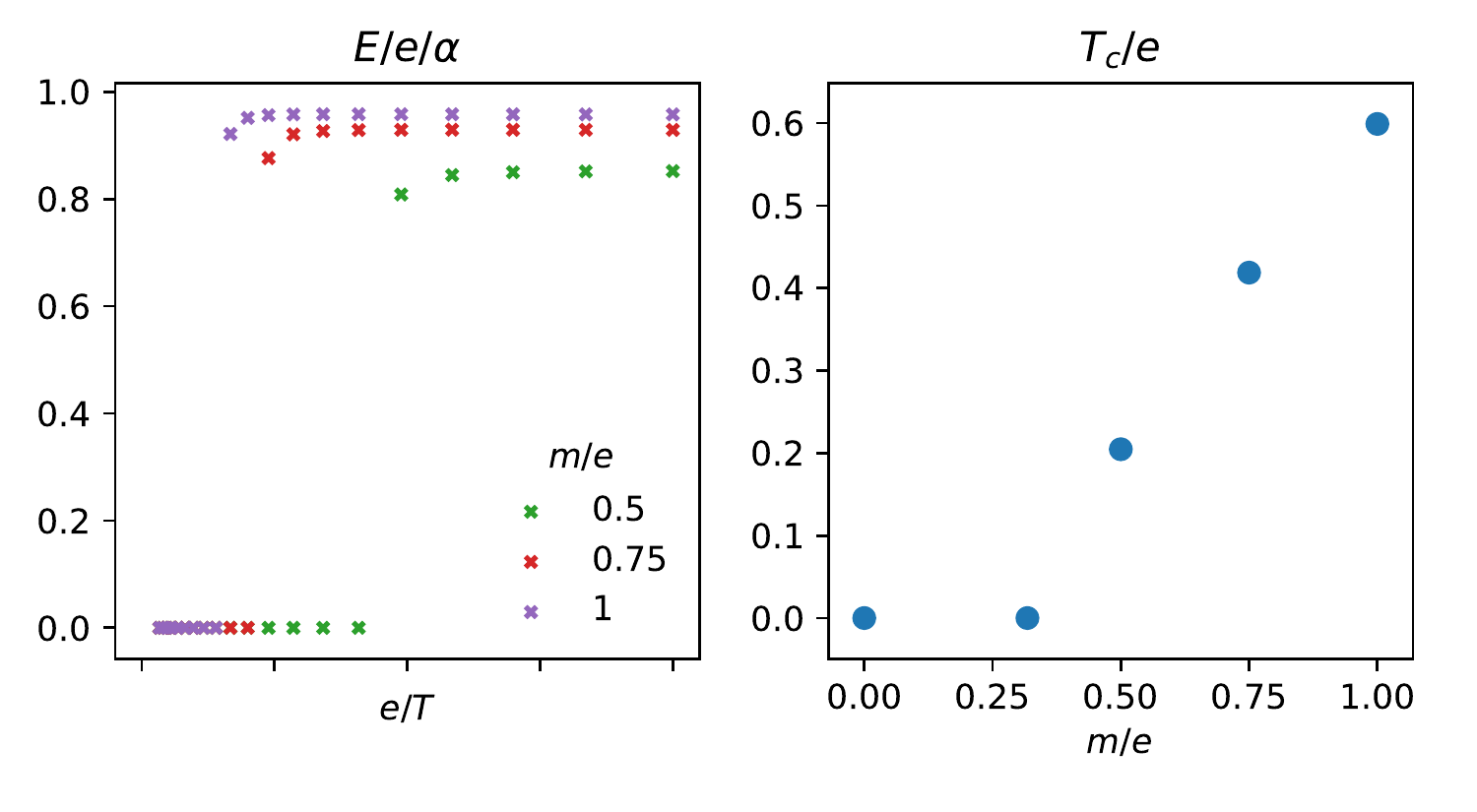}
    \caption{For $\theta=\pi$, we observe that the symmetry of the effective potential is not restored below a critical temperature $T_c(m/e)$. The left plot shows the temperature dependence of the electric field for different values of $m/e$, suggesting a discontinuous phase transition, and the right plot shows the dependence of $T_c$ on the ratio $m/e$.}
    \label{fig:finiteTphaseTransition}
\end{figure}

Now turning to the case $\theta=\pi$ we observe a major qualitative difference with the \ac{dmrg} \cite{Buyens:2016ecr} result: For values $m/e>(m/e)_c$ we find that thermal fluctuations are no longer able to restore the symmetry of the
effective potential below a critical temperature
$T_c(m/e)$. The dependence of the critical temperature on the ratio $m/e$ is shown in Fig.~\ref{fig:finiteTphaseTransition} along with the nonvanishing order parameter. This is unexpected since at the scale $k=2\pi T$ the model becomes effectively one dimensional as the nonzero Matsubara frequencies become heavily suppressed. Although, as discussed in Ref.~\cite{Buyens:2016ecr}, neither the Mermin-Wagner theorem nor the Peierls argument apply to the Schwinger model, a one dimensional field theory at $T>0$ corresponds to a quantum-mechanical one-particle problem, where \ac{ssb} is usually assumed to be absent. Also, in Ref.~\cite{Buyens:2016ecr} numerical evidence was found for the restoration of symmetry. 

At finite temperature, the flow is eventually dominated by the vanishing Matsubara frequency $\omega_{n=0}=0$ (quantum-classical crossover). One can then approximate the Matsubara sum by $T \sum_{\omega_n}f(\omega_n)\simeq T f(0)$. The flow equations become one-dimensional and the prefactor $1/\sqrt{Z_k X_k}$ of the loop term in the flow equations (see Appendix~\ref{sec:flow_equations}) is then replaced by 
\begin{equation}
    \frac{\tilde T}{\sqrt{Z_k X_k}} = \frac{T}{Z_k k} \sim k^{\eta_k-1}
\end{equation}
and is suppressed when $\eta_k>1$ ($\tilde T=T/c_kk$ is the dimensionless temperature). This will occur whenever in the quantum part of the flow the system is attracted by the $T=0$ fixed point of the ordered phase characterized by $\lim_{k\to 0}\eta_k=2$. In that case the $T=0$ fixed point remains attractive and the symmetry is not restored. This phenomenon is thus strongly related to the problem of convexity discussed in Sec. \ref{sec:convexity}. As noted in Ref.~\cite{Nandori:2013nda}, a similar difficulty occurs in the FRG approach to the one-dimensional sine-Gordon model. The instantons, which are responsible for the restoration of the symmetry in the one-dimensional sine-Gordon model~\cite{Rajaraman_book}, are therefore not properly taken into account in the derivative expansion although they are accurately described in the two-dimensional case~\cite{DavietDupuis}.

Interestingly, recent studies have shown using symmetry arguments that some specific quantum mechanical systems, like a particle on a ring with magnetic flux going through the ring, can indeed have a degenerate ground state \cite{Komargodski2020SymmetriesQCD2,Komargodski2}. The authors were able to map the free Schwinger model on a torus with no coupled fermions, corresponding to the limit $m/e\rightarrow \infty$, to this particular quantum-mechanical model. Using Lorentz symmetry, this result is then also applicable to finite temperature via a Wick rotation, showing that the free Schwinger model has a degenerate ground state at $T>0$. This was shown to hold also for the gauge field coupled to a matter particle \cite{Komargodski2}. However a requirement of their discussion is that the matter particle must have a charge greater than the fundamental charge, e.g. $q=2e$. Since the model discussed in our present work has the fundamental charge, these arguments are not applicable. We are therefore convinced that the lack of symmetry restoration in the Schwinger model at finite temperature is an artifact of our approach. 

\section{Conclusion}
\label{sec:conclusion}

The nonperturbative functional renormalization group approach has proven to be an efficient method to study the sine-Gordon model. In particular, it gives an accurate estimate of the mass of the soliton and soliton-antisoliton bound state (breather) in the massive phase, which is exactly known thanks to the integrability of the model. It also allows us to quantify the amplitude of quantum fluctuations in agreement with the Lukyanov-Zamolodchikov conjecture~\cite{DavietDupuis}. In this paper, we have shown that the \ac{frg} approach is also a very powerful method, notably at zero temperature, to determine the physical properties of the massive sine-Gordon model, which is the bosonized version of the massive Schwinger model. More generally, the \ac{frg} is very useful to study quantum systems in $d=1+1$ dimensions where solitonlike excitations may play a crucial role~\cite{[{The \ac{frg} has recently been used to study one-dimensional disordered quantum fluids. See, e.g., }] Dupuis20,*Daviet20,*Daviet21}.

By computing the critical exponents, we confirm that the phase transition occurring in the massive Schwinger model for a value $\theta=\pi$ of the vacuum angle belongs to the two-dimensional Ising universality class, as had been shown in several previous studies~\cite{Byrnes:2002nv,Shimizu:2014fsa}. Though the precision of the critical exponents is typical of a second-order derivative expansion (\ac{de}), more accurate results could be obtained by pushing the expansion to higher orders~\cite{Balog19,DePolsi20}. On the other hand, we find that the phase transition occurs for a critical ratio $(m/e)_c=0.318(13)$ between the mass and the charge of the fermions, which is in agreement with the highest precision results available in the literature \cite{Byrnes:2002nv}.

We have also computed the temperature and vacuum angle dependence of various physical quantities: string tension, electric field, chiral density, mass gap of the lowest-lying excitation and entropy density. At zero temperature, our results are in excellent agreement with \ac{dmrg} lattice calculations~\cite{Buyens:2017crb}. The finite temperature results near $\theta=0$ are also in good agreement with the lattice calculations~\cite{Buyens:2016ecr}, though less accurate. At $\theta=\pi$ and for $m/e>(m/e)_c$, we do not observe the restoration of the symmetry as expected for a quantum system in one space dimension at finite temperature. We ascribe this failure to the inability of our limited truncation to properly account for all the effects of instantons in the one-dimensional Schwinger and sine-Gordon models~\cite{Nandori:2013nda}. 

Whether or not topological excitations, in particular when they are responsible for symmetry restoration, can be captured by truncations like the DE is a very interesting issue in many models. There is no doubt that the DE, which yields very accurate estimates of the critical exponents~\cite{Balog19,DePolsi20}, captures the topological excitations of the three-dimensional O(3) model in which the hedgehog singularities are known to be essential~\cite{Motrunich04,Chlebicki21}. The situation is less clear in lower dimensions. Nevertheless, the kinks of the one-dimensional $\varphi^4$ theory~\cite{Rulquin15a} and the vortex excitations (responsible for the Berezinskii-Kosterlitz-Thouless transition) in the two-dimensional O(2) model~\cite{Jakubczyk14} are, at least partially, captured. In the two-dimensional sine-Gordon model solitons and antisolitons, as well as their lowest-lying bound state (breather), are well described~\cite{DavietDupuis}. The possibility to accurately describe the topological excitations and their consequences for the  one-dimensional sine-Gordon model, and therefore the finite-temperature Schwinger model (in $d=1+1$), remains currently an open issue. 


\section{Acknowledgements}

We thank Zohar Komargodski for his valuable insights on the issue of \ac{ssb} at finite temperature, and Bertrand Delamotte for a useful discussion.

The work of S.~F.\ is supported by the Deutsche Forschungsgemeinschaft (DFG, German Research Foundation) under Germany's Excellence Strategy EXC 2181/1 - 390900948 (the Heidelberg STRUCTURES Excellence Cluster), SFB 1225 (ISOQUANT) as well as FL 736/3-1.

\begin{appendix}
    
\section{Regulator Dependence}
\label{sec:pms}

\begin{figure}
    \centering
    \includegraphics[width=\linewidth]{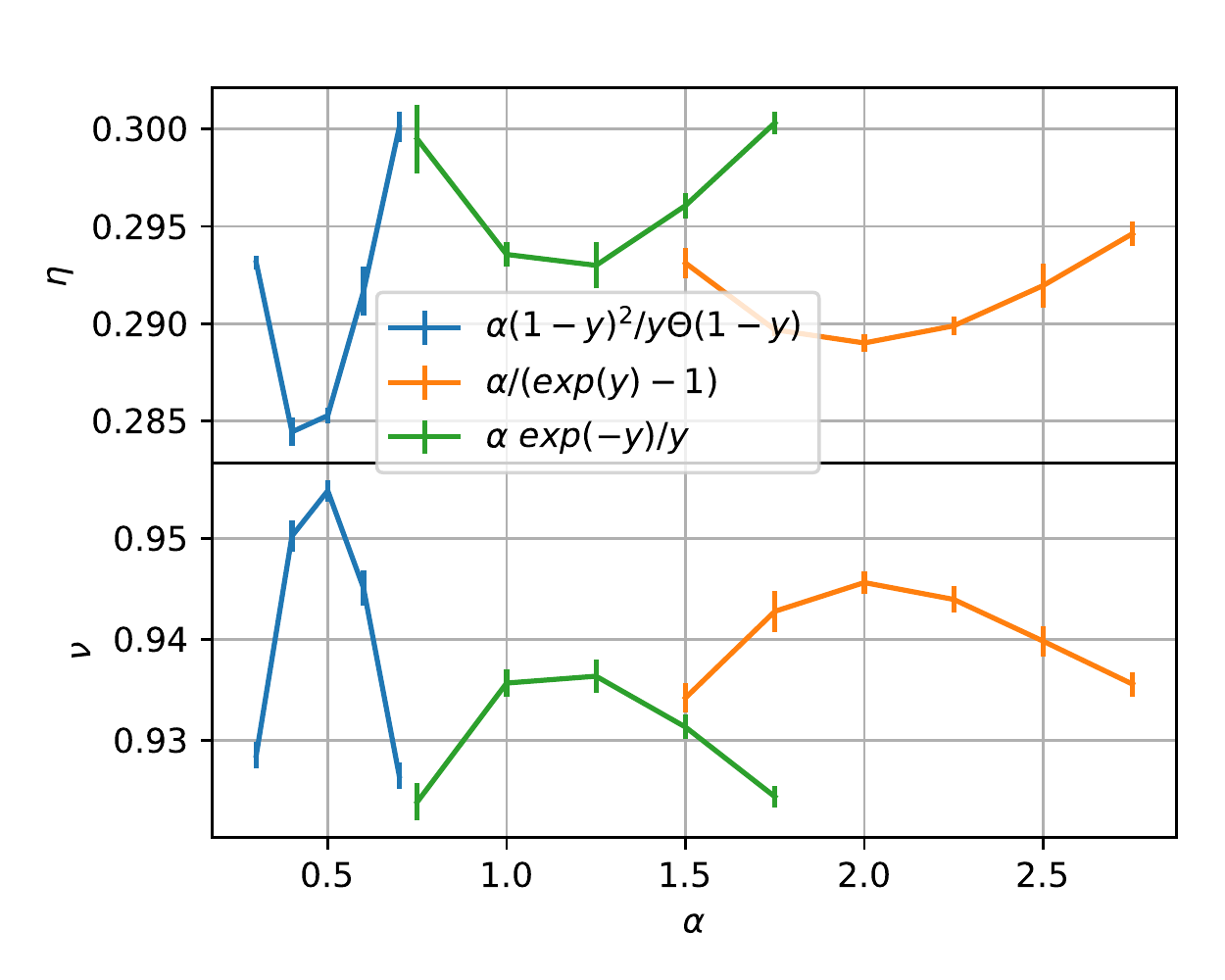}
    \caption{Critical exponents $\eta$ and $\nu$ vs $\alpha$ for the three regulators defined in~(\ref{eq:rdef}).
    }
    \label{fig:pms}
\end{figure}

The critical exponents were calculated using the \acl{pms} and three different regulators,
\begin{equation}
\label{eq:rdef} 
    r(y)=
    \begin{cases}
    \dfrac{\alpha}{e^y-1} , \\
    \alpha \dfrac{(1-y)^2}{y}\Theta(1-y)  , \\
    \alpha\dfrac{e^{-y}}{y} ,
    \end{cases}
\end{equation}
which all have the same limit $r(y)\rightarrow \alpha/y$ for $y\rightarrow 0$ ($\Theta$ denotes the step function). The result is shown in Fig.~\ref{fig:pms}. The first regulator, also defined in Eq. (\ref{eq:regulator}), shows the least sensitivity of the three regulators and was thus chosen with $\alpha = 2$.

\vspace{\columnsep}

\section{Flow equations}
\label{sec:flow_equations}

Defining the length of the system $L$ via $(2\pi)^2\delta(0)=L/T$, the flow equations for $U_k$, $Z_k$ and $X_k$ can be deduced from Eq. (\ref{eq:wetterich}) using the following identities 
\begin{equation}
    U_k(\phi) = \frac{T}{L}\Gamma_k[\phi],
\end{equation}
\begin{equation}
    Z_k(\phi) = \frac{T}{2L}\partial_p^2\frac{\delta^2}{\delta \phi(-p,0)\delta \phi(p,0)}\Gam{2}_k[\phi],
\end{equation}
\begin{equation}
    X_k(\phi) = \frac{T}{2L}\Delta_\omega^2\frac{\delta^2}{\delta \phi(0,-\omega)\delta \phi(0,\omega)}\Gam{2}_k[\phi]
\end{equation}
where $\Delta_\omega$ denotes the finite difference operator 
\begin{equation}
    \Delta_\omega f(\omega) = \frac{1}{2\pi T}(f(\omega+2\pi T)-f(\omega))
\end{equation}
and the right hand sides are to be evaluated at constant field and vanishing frequency and momentum. As discussed in the main text, we find that the projection for $X_k(\phi)$ using a finite difference leads to better results than a continuous derivative.

The flow equations for the dimensionless functions are then given by 
\begin{equation}
\label{eq:floweffpot}
    \partial_t \tilde U_k = (\eta_k-2) \tilde U_k + \frac{1}{2\sqrt{Z_k X_k}} l_{1,0,0}^{0,0},
\end{equation}

\onecolumngrid
\newpage

\begin{equation}
    \begin{aligned}
        \partial_t \tilde Z_k = \eta_k \tilde Z_k +\frac{1}{\sqrt{Z_k X_k}}&\left[
        l_{2,0,1}^{2,0} \tilde{U}_k^{\prime \prime \prime } \tilde{X}_k^{\prime }+\left(2 l_{3,0,0}^{0,0}+2 l_{2,1,0}^{0,1}+l_{2,0,1}^{0,2}\right) \tilde{U}_k^{\prime
        \prime \prime } \tilde{Z}_k^{\prime } -\frac{1}{2} l_{2,0,0}^{0,0} \tilde{Z}_k^{\prime \prime } +\frac{1}{2} l_{2,0,1}^{4,0} \tilde{X}_k^{\prime }{}^2 \right.\\
        &\left.
        +\left(3
        l_{3,0,0}^{0,2}+2 l_{2,1,0}^{0,3}+\frac{1}{2} l_{2,0,1}^{0,4}\right) \tilde{Z}_k^{\prime }{}^2
        +\left(2 l_{3,0,0}^{2,0}+2
        l_{2,1,0}^{2,1}+l_{2,0,1}^{2,2}\right) \tilde{X}_k^{\prime } \tilde{Z}_k^{\prime }
        +\frac{1}{2} l_{2,0,1}^{0,0} \tilde{U}_k^{\prime \prime \prime }{}^2
        \right]
    \end{aligned}
\end{equation}
and
\begin{equation}
    \begin{aligned}
        \partial_t \tilde X_k = \xi_k \tilde X_k +\frac{1}{\sqrt{Z_k X_k}}&\left[t_{2,0,1}^{0,2,0} \tilde{U}_k^{\prime \prime \prime } \tilde{Z}_k^{\prime }+
        \left(2 t_{3,0,0}^{0,0,0}+2 t_{2,1,0}^{1,0,0}+t_{2,0,1}^{2,0,0}+t_{2,0,1}^{0,0,2}\right) \tilde{U}_k^{\prime \prime \prime } \tilde{X}_k^{\prime} -\frac{1}{2} t_{2,0,0}^{0,0,0} \tilde{X}_k^{\prime \prime }+\frac{1}{2} t_{2,0,1}^{0,4,0} \tilde{Z}_k^{\prime }{}^2 \right.\\
        &\left.
        +\left(3 t_{3,0,0}^{2,0,0}+2
        t_{2,1,0}^{3,0,0}+\frac{1}{2} t_{2,0,1}^{4,0,0}+\frac{1}{2} t_{2,0,1}^{0,0,4}+t_{3,0,0}^{0,0,2}+2 t_{2,1,0}^{1,0,2}+\frac{3}{2} t_{2,0,1}^{2,0,2}\right)\tilde{X}_k^{\prime }{}^2
        \right.\\
        &\left.
        +\left(2
        t_{3,0,0}^{0,2,0}+2 t_{2,1,0}^{1,2,0}+t_{2,0,1}^{2,2,0}+t_{2,0,1}^{0,2,2}\right) \tilde{X}_k^{\prime } \tilde{Z}_k^{\prime }
        +\frac{1}{2} t_{2,0,1}^{0,0,0} \tilde{U}_k^{\prime \prime \prime }{}^2
        \right]
    \end{aligned}
\end{equation}
with $t=\ln(k/\Lambda)$. The threshold functions $l_{a,b,c}^{\nu,\mu}=l_{a,b,c}^{\nu,\mu}(\tilde U_k^{\prime\prime}, \tilde Z_k, \tilde X_k, \eta_k,\xi_k)$ are defined as
\begin{equation}
    l_{a,b,c}^{\nu,\mu}=\int_{\tilde p,\tilde \omega} \tilde \omega^\nu \tilde p^\mu \tilde G^a (\partial_{\tilde p}  \tilde G)^b (\partial_{\tilde p}^2 \tilde G)^c \frac{\dot R_k}{Z_k k^2}
\end{equation}
and, analogously, $t_{a,b,c}^{\nu,\mu,f}=t_{a,b,c}^{\nu,\mu,f}(\tilde U_k^{\prime\prime}, \tilde Z_k, \tilde X_k, \eta_k,\xi_k)$ with
\begin{equation}
    t_{a,b,c}^{\nu,\mu,f}=(2\pi \tilde T)^f \int_{\tilde p,\tilde \omega} \tilde \omega^\nu \tilde p^\mu \tilde G^a (\Delta_{\tilde \omega} \tilde G)^b (\Delta_{\tilde \omega}^2 \tilde G)^c \frac{\dot R_k}{Z_k k^2} ,
\end{equation}
where $\tilde p=p/k$ and $\tilde\omega=\omega/c_kk$.
The dimensionless propagator and its derivatives read
\begin{equation}
    \tilde G=\frac{1}{\tilde Z_k \tilde p^2 + \tilde X_k \tilde \omega^2 + \tilde U_k^{\prime\prime} + y r },
\end{equation}
\begin{equation}
    \partial_{\tilde p} \tilde G = -\tilde G^2 (\tilde Z_k + r + y r^\prime),
\hspace{1cm}
    \partial_{\tilde p}^2 \tilde G = 2\tilde G^3 (\tilde Z_k + r + y r^\prime)^2 -\tilde G^2 (2r^\prime + yr^{\prime\prime}) ,
\end{equation}
where $y=\tilde p^2+\tilde\omega^2$.
In the zero temperature limit $T\rightarrow 0$, the finite difference operator becomes continuous $\Delta_\omega\rightarrow\partial_\omega$ and the threshold functions reduce to $t_{a,b,c}^{n,m,0} \rightarrow l_{a,b,c}^{n,m}$. The Euclidean $SO(2)$ symmetry is then restored, $X_k(\phi)=Z_k(\phi)$, so that the two flow equations for $\tilde Z_k$ and $\tilde X_k$ become identical, 
\begin{equation}
\label{eq:flowwfren}
    \begin{aligned}
        \partial_t \tilde Z_k = \eta_k \tilde Z_k +\frac{1}{Z_k}&\left[
        \left(2 l_{3,0,0}^{0,0}+2 l_{2,1,0}^{0,1}+l_{2,0,1}^{0,2}+l_{2,0,1}^{2,0}\right) \tilde{U}_k^{\prime \prime \prime } \tilde{Z}_k^{\prime }-\frac{1}{2} l_{2,0,0}^{0,0} \tilde{Z}_k^{\prime \prime }+\frac{1}{2}
        l_{2,0,1}^{0,0} \tilde{U}_k^{\prime \prime \prime }{}^2 \right.\\
        &\left.
        +\left(3 l_{3,0,0}^{0,2}+2 l_{2,1,0}^{0,3}+\frac{1}{2} l_{2,0,1}^{0,4}+2 l_{3,0,0}^{2,0}+2
        l_{2,1,0}^{2,1}+l_{2,0,1}^{2,2}+\frac{1}{2} l_{2,0,1}^{4,0}\right) \tilde{Z}_k^{\prime }{}^2
        \right] ,
    \end{aligned}
\end{equation}
and agree with the equation found in \cite{DavietDupuis} after a partial integration. 

\section{Flow equations for the chiral condensate}
\label{sec:flowchiral}

The chiral condensate can be obtained by integrating the flow equations for the derivatives of the functions $\tilde U_k$, $\tilde Z_k$, and $\tilde X_k$ with respect to the fermion mass $m$. These flow equations can be obtained from the flow equations of the effective potential and the wave function renormalizations,
\begin{equation}
    \begin{aligned}
    \partial_t \partial_m \tilde U_k = (\eta_k-2) \partial_m\tilde U_k + \frac{1}{2\sqrt{Z_k X_k}} \partial_m l_{1,0,0}^{0,0} ,
    \end{aligned}
\end{equation}
\begin{equation}
    \begin{aligned}
        \partial_t \partial_m  \tilde Z_k = \eta_k \partial_m \tilde Z_k +\frac{1}{\sqrt{Z_k X_k}}&\left[
        (\partial_m l_{2,0,1}^{2,0}) \tilde{U}_k^{\prime \prime \prime } \tilde{X}_k^{\prime }+l_{2,0,1}^{2,0}(\partial_m \tilde{U}_k^{\prime \prime \prime }) \tilde{X}_k^{\prime }+l_{2,0,1}^{2,0} \tilde{U}_k^{\prime \prime \prime } (\partial_m\tilde{X}_k^{\prime })\right.\\
        &+\left.\left(2 \partial_m l_{3,0,0}^{0,0}+2 \partial_m l_{2,1,0}^{0,1}+\partial_m l_{2,0,1}^{0,2}\right) \tilde{U}_k^{\prime
        \prime \prime } \tilde{Z}_k^{\prime }\right.\\
        &+\left.\left(2 l_{3,0,0}^{0,0}+2 l_{2,1,0}^{0,1}+l_{2,0,1}^{0,2}\right) (\partial_m(\tilde{U}_k^{\prime
        \prime \prime }) \tilde{Z}_k^{\prime } + \tilde{U}_k^{\prime
        \prime \prime } (\partial_m\tilde{Z}_k^{\prime }))\right.\\
        &+\left.\frac{1}{2} (\partial_m l_{2,0,1}^{4,0}) \tilde{X}_k^{\prime }{}^2+ l_{2,0,1}^{4,0} (\partial_m \tilde{X}_k^{\prime })\tilde{X}_k^{\prime }\right.\\
        &\left.+\frac{1}{2} (\partial_m l_{2,0,1}^{0,0}) \tilde{U}_k^{\prime \prime \prime }{}^2 + l_{2,0,1}^{0,0}(\partial_m \tilde{U}_k^{\prime \prime \prime }) \tilde{U}_k^{\prime \prime \prime } \right.\\
        &\left.
        +\left(2\partial_m l_{3,0,0}^{2,0}+2
        \partial_m l_{2,1,0}^{2,1}+\partial_m l_{2,0,1}^{2,2}\right) \tilde{X}_k^{\prime } \tilde{Z}_k^{\prime }\right.\\
        &+\left.\left(2 l_{3,0,0}^{2,0}+2
        l_{2,1,0}^{2,1}+l_{2,0,1}^{2,2}\right) ((\partial_m\tilde{X}_k^{\prime }) \tilde{Z}_k^{\prime }+\tilde{X}_k^{\prime }(\partial_m \tilde{Z}_k^{\prime }))\right.\\
        &\left.-\frac{1}{2} (\partial_m l_{2,0,0}^{0,0}) \tilde{Z}_k^{\prime \prime } -\frac{1}{2} l_{2,0,0}^{0,0}(\partial_m \tilde{Z}_k^{\prime \prime })\right.\\
        &\left.
        +\left(3
        \partial_m l_{3,0,0}^{0,2}+2 \partial_m l_{2,1,0}^{0,3}+\frac{1}{2} \partial_m l_{2,0,1}^{0,4}\right) \tilde{Z}_k^{\prime }{}^2
        \right.\\
        &\left.
        +2\left(3
        l_{3,0,0}^{0,2}+2 l_{2,1,0}^{0,3}+\frac{1}{2} l_{2,0,1}^{0,4}\right) (\partial_m \tilde{Z}_k^{\prime })\tilde{Z}_k^{\prime }
        \right] 
    \end{aligned}
\end{equation}
and
\begin{equation}
    \begin{aligned}
        \partial_t\partial_m \tilde X_k = \xi_k\partial_m \tilde X_k +\frac{1}{\sqrt{Z_k X_k}}&\left[
        (\partial_m t_{2,0,1}^{0,2,0}) \tilde{U}_k^{\prime \prime \prime } \tilde{Z}_k^{\prime } +t_{2,0,1}^{0,2,0}(\partial_m \tilde{U}_k^{\prime \prime \prime }) \tilde{Z}_k^{\prime } +t_{2,0,1}^{0,2,0} \tilde{U}_k^{\prime \prime \prime } (\partial_m\tilde{Z}_k^{\prime })\right.\\
        &+\left.\left(2 \partial_mt_{3,0,0}^{0,0,0}+2 \partial_mt_{2,1,0}^{1,0,0}+\partial_mt_{2,0,1}^{2,0,0}+\partial_mt_{2,0,1}^{0,0,2}\right) \tilde{U}_k^{\prime \prime \prime } \tilde{X}_k^{\prime}\right.\\
        &+\left.\left(2 t_{3,0,0}^{0,0,0}+2 t_{2,1,0}^{1,0,0}+t_{2,0,1}^{2,0,0}+t_{2,0,1}^{0,0,2}\right) ((\partial_m\tilde{U}_k^{\prime \prime \prime }) \tilde{X}_k^{\prime}+\tilde{U}_k^{\prime \prime \prime } (\partial_m\tilde{X}_k^{\prime}))\right.\\
        &\left.
        +\frac{1}{2} (\partial_m t_{2,0,1}^{0,0,0}) \tilde{U}_k^{\prime \prime \prime }{}^2 
        + t_{2,0,1}^{0,0,0} (\partial_m\tilde{U}_k^{\prime \prime \prime })\tilde{U}_k^{\prime \prime \prime } \right.\\
        &\left.+\left(2
        \partial_mt_{3,0,0}^{0,2,0}+2 \partial_mt_{2,1,0}^{1,2,0}+\partial_mt_{2,0,1}^{2,2,0}+\partial_mt_{2,0,1}^{0,2,2}\right) \tilde{X}_k^{\prime } \tilde{Z}_k^{\prime }\right.\\
        &\left.+\left(2
        t_{3,0,0}^{0,2,0}+2 t_{2,1,0}^{1,2,0}+t_{2,0,1}^{2,2,0}+t_{2,0,1}^{0,2,2}\right) ((\partial_m\tilde{X}_k^{\prime }) \tilde{Z}_k^{\prime }+\tilde{X}_k^{\prime } (\partial_m\tilde{Z}_k^{\prime }))\right.\\
        &\left.
        +\left(3 \partial_mt_{3,0,0}^{2,0,0}+2
        \partial_mt_{2,1,0}^{3,0,0}+\frac{1}{2} \partial_mt_{2,0,1}^{4,0,0}+\frac{1}{2} \partial_mt_{2,0,1}^{0,0,4}+\partial_mt_{3,0,0}^{0,0,2} \right.\right.\\
        &\left.\left.+2 \partial_mt_{2,1,0}^{1,0,2}+\frac{3}{2} \partial_mt_{2,0,1}^{2,0,2}\right)\tilde{X}_k^{\prime }{}^2\right.\\
        &\left.
        +2\left(3 t_{3,0,0}^{2,0,0}+2
        t_{2,1,0}^{3,0,0}+\frac{1}{2} t_{2,0,1}^{4,0,0}+\frac{1}{2} t_{2,0,1}^{0,0,4}+t_{3,0,0}^{0,0,2}+2 t_{2,1,0}^{1,0,2}+\frac{3}{2} t_{2,0,1}^{2,0,2}\right)(\partial_m\tilde{X}_k^{\prime })\tilde{X}_k^{\prime }\right.\\
        &\left.-\frac{1}{2} (\partial_mt_{2,0,0}^{0,0,0}) \tilde{X}_k^{\prime \prime }-\frac{1}{2} t_{2,0,0}^{0,0,0} (\partial_m\tilde{X}_k^{\prime \prime })+\frac{1}{2} \partial_m t_{2,0,1}^{0,4,0} \tilde{Z}_k^{\prime }{}^2+\frac{1}{2} t_{2,0,1}^{0,4,0} (\partial_m\tilde{Z}_k^{\prime })\tilde{Z}_k^{\prime }\right].
    \end{aligned}
\end{equation}
At zero temperature the equations for the derivatives of the wave function renormalizations reduce to
\begin{equation}
    \begin{aligned}
        \partial_t \partial_m\tilde Z_k = \eta_k \tilde Z_k +\frac{1}{Z_k}&\left[
        \frac{1}{2}
        (\partial_m l_{2,0,1}^{0,0}) \tilde{U}_k^{\prime \prime \prime }{}^2+
        l_{2,0,1}^{0,0} (\partial_m\tilde{U}_k^{\prime \prime \prime })\tilde{U}_k^{\prime \prime \prime } -\frac{1}{2} (\partial_ml_{2,0,0}^{0,0}) \tilde{Z}_k^{\prime \prime }-\frac{1}{2} l_{2,0,0}^{0,0} (\partial_m\tilde{Z}_k^{\prime \prime })\right.\\
        &\left.+ \left(2 \partial_ml_{3,0,0}^{0,0}+2 \partial_ml_{2,1,0}^{0,1}+\partial_ml_{2,0,1}^{0,2}+\partial_ml_{2,0,1}^{2,0}\right) \tilde{U}_k^{\prime \prime \prime } \tilde{Z}_k^{\prime }\right.\\
        &\left.+ \left(2 l_{3,0,0}^{0,0}+2 l_{2,1,0}^{0,1}+l_{2,0,1}^{0,2}+l_{2,0,1}^{2,0}\right) ((\partial_m\tilde{U}_k^{\prime \prime \prime }) \tilde{Z}_k^{\prime }+\tilde{U}_k^{\prime \prime \prime }(\partial_m \tilde{Z}_k^{\prime }))\right.\\
        &\left.
        +\left(3 \partial_m l_{3,0,0}^{0,2}+2 \partial_m l_{2,1,0}^{0,3}+\frac{1}{2} \partial_m l_{2,0,1}^{0,4}+2 \partial_m l_{3,0,0}^{2,0}+2
        \partial_m l_{2,1,0}^{2,1}\right.\right.\\
        &\left. \left.+\partial_m l_{2,0,1}^{2,2}+\frac{1}{2} \partial_m l_{2,0,1}^{4,0}\right) \tilde{Z}_k^{\prime }{}^2\right.\\
        &\left.
        +2\left(3 l_{3,0,0}^{0,2}+2 l_{2,1,0}^{0,3}+\frac{1}{2} l_{2,0,1}^{0,4}+2 l_{3,0,0}^{2,0}+2
        l_{2,1,0}^{2,1}+l_{2,0,1}^{2,2}+\frac{1}{2} l_{2,0,1}^{4,0}\right) (\partial_m \tilde{Z}_k^{\prime })\tilde{Z}_k^{\prime }
        \right].
    \end{aligned}
\end{equation}
The derivatives of the threshold functions are
\begin{equation}
\begin{aligned}
    \partial_m l_{a,b,c}^{\nu,\mu}&=\int_{\tilde p,\tilde \omega} \tilde \omega^\nu \tilde p^\mu \tilde G^a (\partial_{\tilde p}  \tilde G)^b (\partial_{\tilde p}^2 \tilde G)^c \frac{\dot R_k}{Z_k k^2}\\
    &\times\left(a\tilde G^{-1}\partial_m \tilde G+b(\partial_{\tilde p}\tilde G)^{-1}\partial_{\tilde p}\partial_m \tilde G+c(\partial_{\tilde p}^2\tilde G)^{-1}\partial_{\tilde p}^2\partial_m \tilde G\right)
\end{aligned}
\end{equation}
and
\begin{equation}
\begin{aligned}
    \partial_m t_{a,b,c}^{\nu,\mu,f}&=(2\pi \tilde T)^f \int_{\tilde p,\tilde \omega} \tilde \omega^\nu \tilde p^\mu \tilde G^{a-1} (\Delta_{\tilde \omega} \tilde G)^b (\Delta_{\tilde \omega}^2 \tilde G)^c \frac{\dot R_k}{Z_k k^2}\\
    &\times\left(a\tilde G^{-1}\partial_m \tilde G+b(\Delta_{\tilde \omega}\tilde G)^{-1}\Delta_{\tilde \omega}\partial_m \tilde G+c(\Delta_{\tilde \omega}^2\tilde G)^{-1}\Delta_{\tilde \omega}^2\partial_m \tilde G\right)
\end{aligned}
\end{equation}
\twocolumngrid
where the derivative of the propagator is given as
\begin{equation}
    \partial_m \tilde G = -\tilde G^2(\partial_m \tilde Z_k \tilde p^2 + \partial_m \tilde X_k \tilde \omega^2 + \partial_m \tilde U_k^{\prime\prime}).
\end{equation}
The initial conditions for the various functions are
\begin{equation}
\begin{aligned}
    &\partial_m Z_{k_\text{in}}(\phi) = 0 , \quad \partial_m X_{k_\text{in}}(\phi) = 0 , \\ 
    &\partial_m U_{k_\text{in}}(\phi)=- \frac{e^\gamma}{2\pi} \Lambda \cos(\sqrt{4\pi}\phi + \theta).
\end{aligned}
\end{equation}

\begin{figure}[H]
    \centering
    \includegraphics[width=0.7\linewidth]{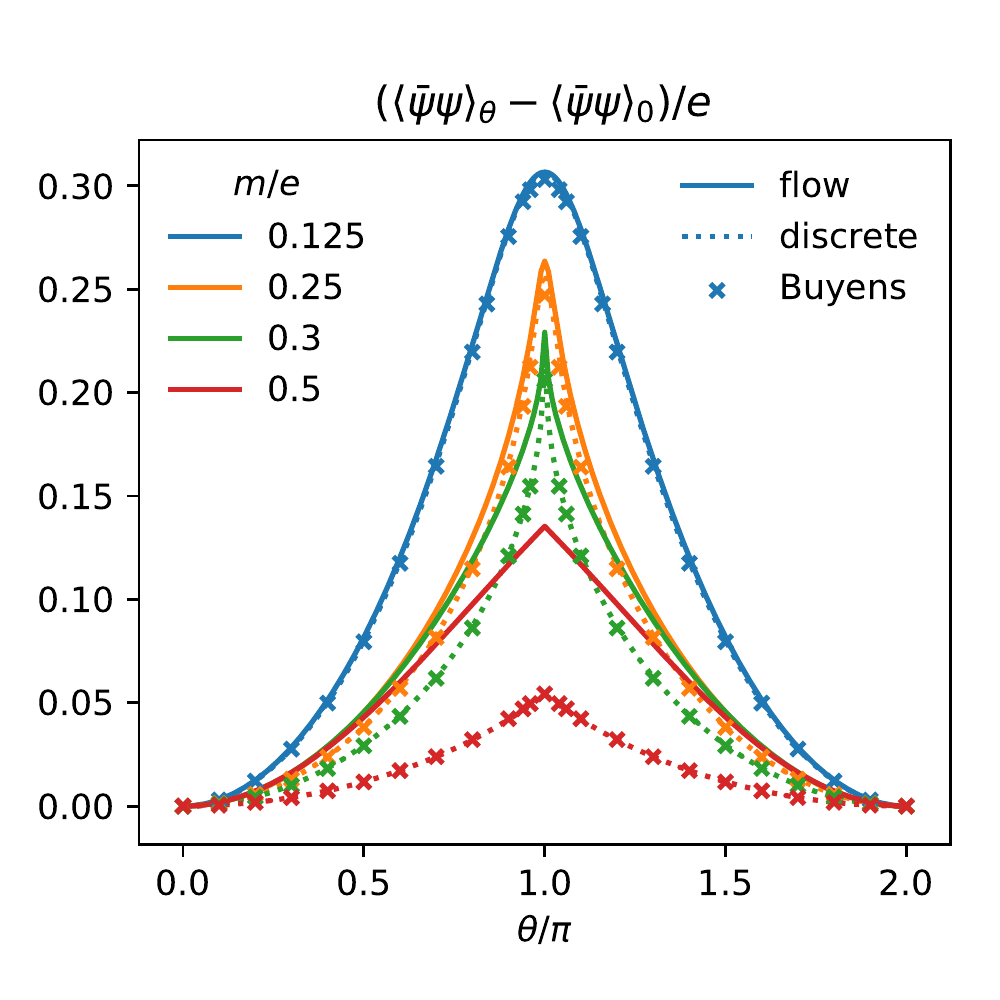}
    \caption{
    Chiral density obtained by integrating the flow equation of $\partial_m U_\theta$ (solid lines), or $U_\theta$ and computing the $m$ derivative numerically by taking a discrete derivative (dotted lines). Only the latter compares well with the \ac{dmrg} result from Buyens \textit{et al.}~\cite{Buyens:2016ecr} for higher values of $m/e$.
    }
    \label{fig:chiraldensity}
\end{figure}

The expectation value of the chiral condensate is then obtained by using Eq.~(\ref{eq:chiraldensity}). The results are compared with those obtained through a discrete derivative in Fig.~\ref{fig:chiraldensity}. Contrary to the discrete-derivative method the flow-equation approach works only for small values of $e/m$. The reason is probably similar to the one invoked in Sec.~\ref{sec:ireval} for the explanation of the deviation of the mass gap from the \ac{dmrg} results.


\section{Numerical implementation}

The hard \ac{uv} cutoff is implemented by limiting the integration or summation boundary by $y<\min(\Lambda^2/k^2,25)$ where $y=(p^2+\omega_n^2/c_k^2)/k^2$. The additional cutoff at $y_\text{max}=25$ at which the integral is already suppressed by a factor of $e^{-25}$ is considered for numerical reasons. If the number of Matsubara frequencies falling into this interval is greater than $N_\text{max}=30$, the flow is approximated by the zero temperature equations. $N_\text{max}=30$ was determined by requiring that the error of the quantity $\eta_k-\xi_k$ due to this approximation is indistinguishable from numeric noise. This quantity is vanishing at zero temperature and therefore an indicator of the importance of thermal fluctuations. The integrals are implemented using the \ac{gsl}. The derivatives of the functions $\tilde U_k(\phi)$, $\tilde Z_k(\phi)$, $\tilde X_k(\phi)$ and their derivatives with respect to $m$ are evaluated in Fourier space, exploiting their periodicity. This is implemented with the \ac{gsl} \acl{fft}. The variable $\phi$ was discretized and a grid size $N=128$ for the various functions was found to be sufficient. Finally, the flow equations are integrated using a \acl{rk4} algorithm. 
    
\end{appendix}
    
\bibliography{references}

\begin{thebibliography}{67}%
\makeatletter
\providecommand \@ifxundefined [1]{%
 \@ifx{#1\undefined}
}%
\providecommand \@ifnum [1]{%
 \ifnum #1\expandafter \@firstoftwo
 \else \expandafter \@secondoftwo
 \fi
}%
\providecommand \@ifx [1]{%
 \ifx #1\expandafter \@firstoftwo
 \else \expandafter \@secondoftwo
 \fi
}%
\providecommand \natexlab [1]{#1}%
\providecommand \enquote  [1]{``#1''}%
\providecommand \bibnamefont  [1]{#1}%
\providecommand \bibfnamefont [1]{#1}%
\providecommand \citenamefont [1]{#1}%
\providecommand \href@noop [0]{\@secondoftwo}%
\providecommand \href [0]{\begingroup \@sanitize@url \@href}%
\providecommand \@href[1]{\@@startlink{#1}\@@href}%
\providecommand \@@href[1]{\endgroup#1\@@endlink}%
\providecommand \@sanitize@url [0]{\catcode `\\12\catcode `\$12\catcode
  `\&12\catcode `\#12\catcode `\^12\catcode `\_12\catcode `\%12\relax}%
\providecommand \@@startlink[1]{}%
\providecommand \@@endlink[0]{}%
\providecommand \url  [0]{\begingroup\@sanitize@url \@url }%
\providecommand \@url [1]{\endgroup\@href {#1}{\urlprefix }}%
\providecommand \urlprefix  [0]{URL }%
\providecommand \Eprint [0]{\href }%
\providecommand \doibase [0]{https://doi.org/}%
\providecommand \selectlanguage [0]{\@gobble}%
\providecommand \bibinfo  [0]{\@secondoftwo}%
\providecommand \bibfield  [0]{\@secondoftwo}%
\providecommand \translation [1]{[#1]}%
\providecommand \BibitemOpen [0]{}%
\providecommand \bibitemStop [0]{}%
\providecommand \bibitemNoStop [0]{.\EOS\space}%
\providecommand \EOS [0]{\spacefactor3000\relax}%
\providecommand \BibitemShut  [1]{\csname bibitem#1\endcsname}%
\let\auto@bib@innerbib\@empty
\bibitem [{\citenamefont {Kadanoff}(1966)}]{kadanoff1966}%
  \BibitemOpen
  \bibfield  {author} {\bibinfo {author} {\bibfnamefont {L.~P.}\ \bibnamefont
  {Kadanoff}},\ }\bibfield  {title} {\bibinfo {title} {Scaling laws for ising
  models near ${T}_{c}$},\ }\href
  {https://doi.org/10.1103/PhysicsPhysiqueFizika.2.263} {\bibfield  {journal}
  {\bibinfo  {journal} {Phys. Phys. Fiz.}\ }\textbf {\bibinfo {volume} {2}},\
  \bibinfo {pages} {263} (\bibinfo {year} {1966})}\BibitemShut {NoStop}%
\bibitem [{\citenamefont {Wilson}(1971{\natexlab{a}})}]{Wilson:1971bg}%
  \BibitemOpen
  \bibfield  {author} {\bibinfo {author} {\bibfnamefont {K.~G.}\ \bibnamefont
  {Wilson}},\ }\bibfield  {title} {\bibinfo {title} {{Renormalization group and
  critical phenomena. 1. Renormalization group and the Kadanoff scaling
  picture}},\ }\href {https://doi.org/10.1103/PhysRevB.4.3174} {\bibfield
  {journal} {\bibinfo  {journal} {Phys. Rev. B}\ }\textbf {\bibinfo {volume}
  {4}},\ \bibinfo {pages} {3174} (\bibinfo {year}
  {1971}{\natexlab{a}})}\BibitemShut {NoStop}%
\bibitem [{\citenamefont {Wilson}(1971{\natexlab{b}})}]{Wilson:1971dh}%
  \BibitemOpen
  \bibfield  {author} {\bibinfo {author} {\bibfnamefont {K.~G.}\ \bibnamefont
  {Wilson}},\ }\bibfield  {title} {\bibinfo {title} {{Renormalization group and
  critical phenomena. 2. Phase space cell analysis of critical behavior}},\
  }\href {https://doi.org/10.1103/PhysRevB.4.3184} {\bibfield  {journal}
  {\bibinfo  {journal} {Phys. Rev. B}\ }\textbf {\bibinfo {volume} {4}},\
  \bibinfo {pages} {3184} (\bibinfo {year} {1971}{\natexlab{b}})}\BibitemShut
  {NoStop}%
\bibitem [{\citenamefont {Wilson}\ and\ \citenamefont
  {Kogut}(1974)}]{Wilson74}%
  \BibitemOpen
  \bibfield  {author} {\bibinfo {author} {\bibfnamefont {K.~G.}\ \bibnamefont
  {Wilson}}\ and\ \bibinfo {author} {\bibfnamefont {J.~B.}\ \bibnamefont
  {Kogut}},\ }\bibfield  {title} {\bibinfo {title} {{The renormalization group
  and the $\epsilon$ expansion}},\ }\href
  {https://doi.org/doi:10.1016/0370-1573(74)90023-4} {\bibfield  {journal}
  {\bibinfo  {journal} {Phys. Rep.}\ }\textbf {\bibinfo {volume} {12}},\
  \bibinfo {pages} {75} (\bibinfo {year} {1974})}\BibitemShut {NoStop}%
\bibitem [{\citenamefont {Wetterich}(1993)}]{Wetterich:1992yh}%
  \BibitemOpen
  \bibfield  {author} {\bibinfo {author} {\bibfnamefont {C.}~\bibnamefont
  {Wetterich}},\ }\bibfield  {title} {\bibinfo {title} {{Exact evolution
  equation for the effective potential}},\ }\href
  {https://doi.org/10.1016/0370-2693(93)90726-X} {\bibfield  {journal}
  {\bibinfo  {journal} {Phys. Lett. B}\ }\textbf {\bibinfo {volume} {301}},\
  \bibinfo {pages} {90} (\bibinfo {year} {1993})}\BibitemShut {NoStop}%
\bibitem [{\citenamefont {Ellwanger}(1994)}]{Ellwanger:1993mw}%
  \BibitemOpen
  \bibfield  {author} {\bibinfo {author} {\bibfnamefont {U.}~\bibnamefont
  {Ellwanger}},\ }\bibfield  {title} {\bibinfo {title} {Flow equations for $n$
  point functions and bound states},\ }\href
  {https://doi.org/10.1007/BF01555911} {\bibfield  {journal} {\bibinfo
  {journal} {Z. Phys. C}\ }\textbf {\bibinfo {volume} {62}},\ \bibinfo {pages}
  {503} (\bibinfo {year} {1994})}\BibitemShut {NoStop}%
\bibitem [{\citenamefont {Morris}(1994)}]{Morris:1993qb}%
  \BibitemOpen
  \bibfield  {author} {\bibinfo {author} {\bibfnamefont {T.~R.}\ \bibnamefont
  {Morris}},\ }\bibfield  {title} {\bibinfo {title} {{The Exact renormalization
  group and approximate solutions}},\ }\href
  {https://doi.org/10.1142/S0217751X94000972} {\bibfield  {journal} {\bibinfo
  {journal} {Int. J. Mod. Phys. A}\ }\textbf {\bibinfo {volume} {9}},\ \bibinfo
  {pages} {2411} (\bibinfo {year} {1994})}\BibitemShut {NoStop}%
\bibitem [{\citenamefont {Berges}\ \emph {et~al.}(2002)\citenamefont {Berges},
  \citenamefont {Tetradis},\ and\ \citenamefont
  {Wetterich}}]{WetterichTetradis}%
  \BibitemOpen
  \bibfield  {author} {\bibinfo {author} {\bibfnamefont {J.}~\bibnamefont
  {Berges}}, \bibinfo {author} {\bibfnamefont {N.}~\bibnamefont {Tetradis}},\
  and\ \bibinfo {author} {\bibfnamefont {C.}~\bibnamefont {Wetterich}},\
  }\bibfield  {title} {\bibinfo {title} {{Non-perturbative renormalization flow
  in quantum field theory and statistical physics}},\ }\href
  {https://doi.org/10.1016/S0370-1573(01)00098-9} {\bibfield  {journal}
  {\bibinfo  {journal} {Phys. Rep.}\ }\textbf {\bibinfo {volume} {363}},\
  \bibinfo {pages} {223} (\bibinfo {year} {2002})}\BibitemShut {NoStop}%
\bibitem [{\citenamefont {Delamotte}(2012)}]{Delamotte12}%
  \BibitemOpen
  \bibfield  {author} {\bibinfo {author} {\bibfnamefont {B.}~\bibnamefont
  {Delamotte}},\ }\bibfield  {title} {\bibinfo {title} {{An Introduction to the
  Nonperturbative Renormalization Group}},\ }in\ \href@noop {} {\emph {\bibinfo
  {booktitle} {{Renormalization Group and Effective Field Theory Approaches to
  Many-Body Systems}}}},\ \bibinfo {series} {Lecture Notes in Physics}, Vol.\
  \bibinfo {volume} {852},\ \bibinfo {editor} {edited by\ \bibinfo {editor}
  {\bibfnamefont {A.}~\bibnamefont {Schwenk}}\ and\ \bibinfo {editor}
  {\bibfnamefont {J.}~\bibnamefont {Polonyi}}}\ (\bibinfo  {publisher}
  {Springer Berlin Heidelberg, Berlin},\ \bibinfo {year} {2012})\ pp.\ \bibinfo
  {pages} {49--132}\BibitemShut {NoStop}%
\bibitem [{\citenamefont {Dupuis}\ \emph {et~al.}(2021)\citenamefont {Dupuis},
  \citenamefont {Canet}, \citenamefont {Eichhorn}, \citenamefont {Metzner},
  \citenamefont {Pawlowski}, \citenamefont {Tissier},\ and\ \citenamefont
  {Wschebor}}]{Dupuis:2020fhh}%
  \BibitemOpen
  \bibfield  {author} {\bibinfo {author} {\bibfnamefont {N.}~\bibnamefont
  {Dupuis}}, \bibinfo {author} {\bibfnamefont {L.}~\bibnamefont {Canet}},
  \bibinfo {author} {\bibfnamefont {A.}~\bibnamefont {Eichhorn}}, \bibinfo
  {author} {\bibfnamefont {W.}~\bibnamefont {Metzner}}, \bibinfo {author}
  {\bibfnamefont {J.}~\bibnamefont {Pawlowski}}, \bibinfo {author}
  {\bibfnamefont {M.}~\bibnamefont {Tissier}},\ and\ \bibinfo {author}
  {\bibfnamefont {N.}~\bibnamefont {Wschebor}},\ }\bibfield  {title} {\bibinfo
  {title} {The nonperturbative functional renormalization group and its
  applications},\ }\href {https://doi.org/10.1016/j.physrep.2021.01.001}
  {\bibfield  {journal} {\bibinfo  {journal} {Phys. Rep.}\ }\textbf {\bibinfo
  {volume} {910}},\ \bibinfo {pages} {1} (\bibinfo {year} {2021})}\BibitemShut
  {NoStop}%
\bibitem [{\citenamefont {Balog}\ \emph {et~al.}(2019)\citenamefont {Balog},
  \citenamefont {Chat\'e}, \citenamefont {Delamotte}, \citenamefont
  {Marohni\'c},\ and\ \citenamefont {Wschebor}}]{Balog19}%
  \BibitemOpen
  \bibfield  {author} {\bibinfo {author} {\bibfnamefont {I.}~\bibnamefont
  {Balog}}, \bibinfo {author} {\bibfnamefont {H.}~\bibnamefont {Chat\'e}},
  \bibinfo {author} {\bibfnamefont {B.}~\bibnamefont {Delamotte}}, \bibinfo
  {author} {\bibfnamefont {M.}~\bibnamefont {Marohni\'c}},\ and\ \bibinfo
  {author} {\bibfnamefont {N.}~\bibnamefont {Wschebor}},\ }\bibfield  {title}
  {\bibinfo {title} {Convergence of nonperturbative approximations to the
  renormalization group},\ }\href
  {https://doi.org/10.1103/PhysRevLett.123.240604} {\bibfield  {journal}
  {\bibinfo  {journal} {Phys. Rev. Lett.}\ }\textbf {\bibinfo {volume} {123}},\
  \bibinfo {pages} {240604} (\bibinfo {year} {2019})}\BibitemShut {NoStop}%
\bibitem [{\citenamefont {De~Polsi}\ \emph {et~al.}(2020)\citenamefont
  {De~Polsi}, \citenamefont {Balog}, \citenamefont {Tissier},\ and\
  \citenamefont {Wschebor}}]{DePolsi20}%
  \BibitemOpen
  \bibfield  {author} {\bibinfo {author} {\bibfnamefont {G.}~\bibnamefont
  {De~Polsi}}, \bibinfo {author} {\bibfnamefont {I.}~\bibnamefont {Balog}},
  \bibinfo {author} {\bibfnamefont {M.}~\bibnamefont {Tissier}},\ and\ \bibinfo
  {author} {\bibfnamefont {N.}~\bibnamefont {Wschebor}},\ }\bibfield  {title}
  {\bibinfo {title} {{Precision calculation of critical exponents in the O($N$)
  universality classes with the nonperturbative renormalization group}},\
  }\href {https://doi.org/10.1103/PhysRevE.101.042113} {\bibfield  {journal}
  {\bibinfo  {journal} {Phys. Rev. E}\ }\textbf {\bibinfo {volume} {101}},\
  \bibinfo {pages} {042113} (\bibinfo {year} {2020})}\BibitemShut {NoStop}%
\bibitem [{\citenamefont {Guida}\ and\ \citenamefont
  {Zinn-Justin}(1998)}]{Guida98}%
  \BibitemOpen
  \bibfield  {author} {\bibinfo {author} {\bibfnamefont {R.}~\bibnamefont
  {Guida}}\ and\ \bibinfo {author} {\bibfnamefont {J.}~\bibnamefont
  {Zinn-Justin}},\ }\bibfield  {title} {\bibinfo {title} {{Critical exponents
  of the $N$-vector model}},\ }\href
  {https://doi.org/doi:10.1088/0305-4470/31/40/006} {\bibfield  {journal}
  {\bibinfo  {journal} {J. Phys. A}\ }\textbf {\bibinfo {volume} {31}},\
  \bibinfo {pages} {8103} (\bibinfo {year} {1998})}\BibitemShut {NoStop}%
\bibitem [{\citenamefont {Kompaniets}\ and\ \citenamefont
  {Panzer}(2017)}]{Kompaniets17}%
  \BibitemOpen
  \bibfield  {author} {\bibinfo {author} {\bibfnamefont {M.~V.}\ \bibnamefont
  {Kompaniets}}\ and\ \bibinfo {author} {\bibfnamefont {E.}~\bibnamefont
  {Panzer}},\ }\bibfield  {title} {\bibinfo {title} {Minimally subtracted
  six-loop renormalization of $o(n)$-symmetric ${\ensuremath{\phi}}^{4}$ theory
  and critical exponents},\ }\href {https://doi.org/10.1103/PhysRevD.96.036016}
  {\bibfield  {journal} {\bibinfo  {journal} {Phys. Rev. D}\ }\textbf {\bibinfo
  {volume} {96}},\ \bibinfo {pages} {036016} (\bibinfo {year}
  {2017})}\BibitemShut {NoStop}%
\bibitem [{\citenamefont {Hasenbusch}(2010)}]{Hasenbusch10}%
  \BibitemOpen
  \bibfield  {author} {\bibinfo {author} {\bibfnamefont {M.}~\bibnamefont
  {Hasenbusch}},\ }\bibfield  {title} {\bibinfo {title} {{Finite size scaling
  study of lattice models in the three-dimensional Ising universality class}},\
  }\href {https://doi.org/10.1103/PhysRevB.82.174433} {\bibfield  {journal}
  {\bibinfo  {journal} {Phys. Rev. B}\ }\textbf {\bibinfo {volume} {82}},\
  \bibinfo {pages} {174433} (\bibinfo {year} {2010})}\BibitemShut {NoStop}%
\bibitem [{\citenamefont {Campostrini}\ \emph {et~al.}(2006)\citenamefont
  {Campostrini}, \citenamefont {Hasenbusch}, \citenamefont {Pelissetto},\ and\
  \citenamefont {Vicari}}]{Campostrini06}%
  \BibitemOpen
  \bibfield  {author} {\bibinfo {author} {\bibfnamefont {M.}~\bibnamefont
  {Campostrini}}, \bibinfo {author} {\bibfnamefont {M.}~\bibnamefont
  {Hasenbusch}}, \bibinfo {author} {\bibfnamefont {A.}~\bibnamefont
  {Pelissetto}},\ and\ \bibinfo {author} {\bibfnamefont {E.}~\bibnamefont
  {Vicari}},\ }\bibfield  {title} {\bibinfo {title} {{Theoretical estimates of
  the critical exponents of the superfluid transition in $^{{4}}\mathrm{{He}}$
  by lattice methods}},\ }\href {https://doi.org/10.1103/PhysRevB.74.144506}
  {\bibfield  {journal} {\bibinfo  {journal} {Phys. Rev. B}\ }\textbf {\bibinfo
  {volume} {74}},\ \bibinfo {pages} {144506} (\bibinfo {year}
  {2006})}\BibitemShut {NoStop}%
\bibitem [{\citenamefont {Campostrini}\ \emph {et~al.}(2002)\citenamefont
  {Campostrini}, \citenamefont {Hasenbusch}, \citenamefont {Pelissetto},
  \citenamefont {Rossi},\ and\ \citenamefont {Vicari}}]{Campostrini02}%
  \BibitemOpen
  \bibfield  {author} {\bibinfo {author} {\bibfnamefont {M.}~\bibnamefont
  {Campostrini}}, \bibinfo {author} {\bibfnamefont {M.}~\bibnamefont
  {Hasenbusch}}, \bibinfo {author} {\bibfnamefont {A.}~\bibnamefont
  {Pelissetto}}, \bibinfo {author} {\bibfnamefont {P.}~\bibnamefont {Rossi}},\
  and\ \bibinfo {author} {\bibfnamefont {E.}~\bibnamefont {Vicari}},\
  }\bibfield  {title} {\bibinfo {title} {{Critical exponents and equation of
  state of the three-dimensional Heisenberg universality class}},\ }\href
  {https://doi.org/10.1103/PhysRevB.65.144520} {\bibfield  {journal} {\bibinfo
  {journal} {Phys. Rev. B}\ }\textbf {\bibinfo {volume} {65}},\ \bibinfo
  {pages} {144520} (\bibinfo {year} {2002})}\BibitemShut {NoStop}%
\bibitem [{\citenamefont {Hasenbusch}(2019)}]{Hasenbusch19}%
  \BibitemOpen
  \bibfield  {author} {\bibinfo {author} {\bibfnamefont {M.}~\bibnamefont
  {Hasenbusch}},\ }\bibfield  {title} {\bibinfo {title} {{Monte Carlo study of
  an improved clock model in three dimensions}},\ }\href
  {https://doi.org/10.1103/PhysRevB.100.224517} {\bibfield  {journal} {\bibinfo
   {journal} {Phys. Rev. B}\ }\textbf {\bibinfo {volume} {100}},\ \bibinfo
  {pages} {224517} (\bibinfo {year} {2019})}\BibitemShut {NoStop}%
\bibitem [{\citenamefont {Clisby}\ and\ \citenamefont
  {D\"unweg}(2016)}]{Clisby16}%
  \BibitemOpen
  \bibfield  {author} {\bibinfo {author} {\bibfnamefont {N.}~\bibnamefont
  {Clisby}}\ and\ \bibinfo {author} {\bibfnamefont {B.}~\bibnamefont
  {D\"unweg}},\ }\bibfield  {title} {\bibinfo {title} {High-precision estimate
  of the hydrodynamic radius for self-avoiding walks},\ }\href
  {https://doi.org/10.1103/PhysRevE.94.052102} {\bibfield  {journal} {\bibinfo
  {journal} {Phys. Rev. E}\ }\textbf {\bibinfo {volume} {94}},\ \bibinfo
  {pages} {052102} (\bibinfo {year} {2016})}\BibitemShut {NoStop}%
\bibitem [{\citenamefont {Clisby}(2017)}]{Clisby17}%
  \BibitemOpen
  \bibfield  {author} {\bibinfo {author} {\bibfnamefont {N.}~\bibnamefont
  {Clisby}},\ }\bibfield  {title} {\bibinfo {title} {{Scale-free Monte Carlo
  method for calculating the critical exponent $\gamma$ of self-avoiding
  walks}},\ }\href {https://doi.org/10.1088/1751-8121/aa7231} {\bibfield
  {journal} {\bibinfo  {journal} {J. Phys. A}\ }\textbf {\bibinfo {volume}
  {50}},\ \bibinfo {pages} {264003} (\bibinfo {year} {2017})}\BibitemShut
  {NoStop}%
\bibitem [{\citenamefont {Kos}\ \emph {et~al.}(2016)\citenamefont {Kos},
  \citenamefont {Poland}, \citenamefont {Simmons-Duffin},\ and\ \citenamefont
  {Vichi}}]{Kos16}%
  \BibitemOpen
  \bibfield  {author} {\bibinfo {author} {\bibfnamefont {F.}~\bibnamefont
  {Kos}}, \bibinfo {author} {\bibfnamefont {D.}~\bibnamefont {Poland}},
  \bibinfo {author} {\bibfnamefont {D.}~\bibnamefont {Simmons-Duffin}},\ and\
  \bibinfo {author} {\bibfnamefont {A.}~\bibnamefont {Vichi}},\ }\bibfield
  {title} {\bibinfo {title} {{Precision islands in the Ising and $O(N)$
  models}},\ }\href {https://doi.org/10.1007/JHEP08(2016)036} {\bibfield
  {journal} {\bibinfo  {journal} {{J. High Energy Phys.}}\ }\textbf {\bibinfo
  {volume} {08}},\ \bibinfo {pages} {036} (\bibinfo {year} {2016})}\BibitemShut
  {NoStop}%
\bibitem [{\citenamefont {Simmons-Duffin}(2017)}]{SimmonsDuffin17}%
  \BibitemOpen
  \bibfield  {author} {\bibinfo {author} {\bibfnamefont {D.}~\bibnamefont
  {Simmons-Duffin}},\ }\bibfield  {title} {\bibinfo {title} {{The lightcone
  bootstrap and the spectrum of the 3d Ising CFT}},\ }\href
  {https://doi.org/10.1007/JHEP03(2017)086} {\bibfield  {journal} {\bibinfo
  {journal} {{J. High Energy Phys.}}\ }\textbf {\bibinfo {volume} {03}},\
  \bibinfo {pages} {086} (\bibinfo {year} {2017})}\BibitemShut {NoStop}%
\bibitem [{\citenamefont {Echeverri}\ \emph {et~al.}(2016)\citenamefont
  {Echeverri}, \citenamefont {von Harling},\ and\ \citenamefont
  {Serone}}]{Echeverri16}%
  \BibitemOpen
  \bibfield  {author} {\bibinfo {author} {\bibfnamefont {A.~C.}\ \bibnamefont
  {Echeverri}}, \bibinfo {author} {\bibfnamefont {B.}~\bibnamefont {von
  Harling}},\ and\ \bibinfo {author} {\bibfnamefont {M.}~\bibnamefont
  {Serone}},\ }\bibfield  {title} {\bibinfo {title} {The effective bootstrap},\
  }\href {https://doi.org/10.1007/JHEP09(2016)097} {\bibfield  {journal}
  {\bibinfo  {journal} {{J. High Energy Phys.}}\ }\textbf {\bibinfo {volume}
  {09}},\ \bibinfo {pages} {097} (\bibinfo {year} {2016})}\BibitemShut
  {NoStop}%
\bibitem [{\citenamefont {Chester}\ \emph {et~al.}(2020)\citenamefont
  {Chester}, \citenamefont {Landry}, \citenamefont {Liu}, \citenamefont
  {Poland}, \citenamefont {Simmons-Duffin}, \citenamefont {Su},\ and\
  \citenamefont {Vichi}}]{Chester20}%
  \BibitemOpen
  \bibfield  {author} {\bibinfo {author} {\bibfnamefont {S.~M.}\ \bibnamefont
  {Chester}}, \bibinfo {author} {\bibfnamefont {W.}~\bibnamefont {Landry}},
  \bibinfo {author} {\bibfnamefont {J.}~\bibnamefont {Liu}}, \bibinfo {author}
  {\bibfnamefont {D.}~\bibnamefont {Poland}}, \bibinfo {author} {\bibfnamefont
  {D.}~\bibnamefont {Simmons-Duffin}}, \bibinfo {author} {\bibfnamefont
  {N.}~\bibnamefont {Su}},\ and\ \bibinfo {author} {\bibfnamefont
  {A.}~\bibnamefont {Vichi}},\ }\bibfield  {title} {\bibinfo {title} {{Carving
  out OPE space and precise O(2) model critical exponents}},\ }\href
  {https://doi.org/10.1007/JHEP06(2020)142} {\bibfield  {journal} {\bibinfo
  {journal} {{J. High Energy Phys.}}\ }\textbf {\bibinfo {volume} {06}},\
  \bibinfo {pages} {142} (\bibinfo {year} {2020})}\BibitemShut {NoStop}%
\bibitem [{\citenamefont {Daviet}\ and\ \citenamefont
  {Dupuis}(2019)}]{DavietDupuis}%
  \BibitemOpen
  \bibfield  {author} {\bibinfo {author} {\bibfnamefont {R.}~\bibnamefont
  {Daviet}}\ and\ \bibinfo {author} {\bibfnamefont {N.}~\bibnamefont
  {Dupuis}},\ }\bibfield  {title} {\bibinfo {title} {Nonperturbative functional
  renormalization-group approach to the sine-gordon model and the
  lukyanov-zamolodchikov conjecture},\ }\href
  {https://doi.org/10.1103/PhysRevLett.122.155301} {\bibfield  {journal}
  {\bibinfo  {journal} {Phys. Rev. Lett.}\ }\textbf {\bibinfo {volume} {122}},\
  \bibinfo {pages} {155301} (\bibinfo {year} {2019})}\BibitemShut {NoStop}%
\bibitem [{\citenamefont {Nandori}\ \emph {et~al.}(2009)\citenamefont
  {Nandori}, \citenamefont {Nagy}, \citenamefont {Sailer},\ and\ \citenamefont
  {Trombettoni}}]{Nandori:2009ad}%
  \BibitemOpen
  \bibfield  {author} {\bibinfo {author} {\bibfnamefont {I.}~\bibnamefont
  {Nandori}}, \bibinfo {author} {\bibfnamefont {S.}~\bibnamefont {Nagy}},
  \bibinfo {author} {\bibfnamefont {K.}~\bibnamefont {Sailer}},\ and\ \bibinfo
  {author} {\bibfnamefont {A.}~\bibnamefont {Trombettoni}},\ }\bibfield
  {title} {\bibinfo {title} {Comparison of renormalization group schemes for
  sine-gordon type models},\ }\href
  {https://doi.org/10.1103/PhysRevD.80.025008} {\bibfield  {journal} {\bibinfo
  {journal} {Phys. Rev. D}\ }\textbf {\bibinfo {volume} {80}},\ \bibinfo
  {pages} {025008} (\bibinfo {year} {2009})}\BibitemShut {NoStop}%
\bibitem [{\citenamefont {Nandori}(2011)}]{Nandori:2010ij}%
  \BibitemOpen
  \bibfield  {author} {\bibinfo {author} {\bibfnamefont {I.}~\bibnamefont
  {Nandori}},\ }\bibfield  {title} {\bibinfo {title} {{Bosonization and
  Functional Renormalization Group Approach in the Framework of QED$_{2}$}},\
  }\href {https://doi.org/10.1103/PhysRevD.84.065024} {\bibfield  {journal}
  {\bibinfo  {journal} {Phys. Rev. D}\ }\textbf {\bibinfo {volume} {84}},\
  \bibinfo {pages} {065024} (\bibinfo {year} {2011})}\BibitemShut {NoStop}%
\bibitem [{\citenamefont {Schwinger}(1962)}]{Schwinger:1962tp}%
  \BibitemOpen
  \bibfield  {author} {\bibinfo {author} {\bibfnamefont {J.~S.}\ \bibnamefont
  {Schwinger}},\ }\bibfield  {title} {\bibinfo {title} {{Gauge Invariance and
  Mass. 2.}},\ }\href {https://doi.org/10.1103/PhysRev.128.2425} {\bibfield
  {journal} {\bibinfo  {journal} {Phys. Rev.}\ }\textbf {\bibinfo {volume}
  {128}},\ \bibinfo {pages} {2425} (\bibinfo {year} {1962})}\BibitemShut
  {NoStop}%
\bibitem [{\citenamefont {Abdalla}\ \emph {et~al.}(2001)\citenamefont
  {Abdalla}, \citenamefont {Abdalla},\ and\ \citenamefont {Rothe}}]{abdallah}%
  \BibitemOpen
  \bibfield  {author} {\bibinfo {author} {\bibfnamefont {E.}~\bibnamefont
  {Abdalla}}, \bibinfo {author} {\bibfnamefont {M.~C.~B.}\ \bibnamefont
  {Abdalla}},\ and\ \bibinfo {author} {\bibfnamefont {K.~D.}\ \bibnamefont
  {Rothe}},\ }\href@noop {} {\emph {\bibinfo {title} {Non-Perturbative Methods
  in 2 Dimensional Quantum Field Theory}}},\ \bibinfo {edition} {2nd}\ ed.\
  (\bibinfo  {publisher} {World Scientific, Singapore},\ \bibinfo {year}
  {2001})\BibitemShut {NoStop}%
\bibitem [{\citenamefont {Lowenstein}\ and\ \citenamefont
  {Swieca}(1971)}]{Lowenstein:1971fc}%
  \BibitemOpen
  \bibfield  {author} {\bibinfo {author} {\bibfnamefont {J.}~\bibnamefont
  {Lowenstein}}\ and\ \bibinfo {author} {\bibfnamefont {J.}~\bibnamefont
  {Swieca}},\ }\bibfield  {title} {\bibinfo {title} {{Quantum electrodynamics
  in two-dimensions}},\ }\href {https://doi.org/10.1016/0003-4916(71)90246-6}
  {\bibfield  {journal} {\bibinfo  {journal} {Ann. Phys. (N.Y.)}\ }\textbf
  {\bibinfo {volume} {68}},\ \bibinfo {pages} {172} (\bibinfo {year}
  {1971})}\BibitemShut {NoStop}%
\bibitem [{\citenamefont {Zinn-Justin}(2002)}]{ZinnJustin:2002ru}%
  \BibitemOpen
  \bibfield  {author} {\bibinfo {author} {\bibfnamefont {J.}~\bibnamefont
  {Zinn-Justin}},\ }\href@noop {} {\emph {\bibinfo {title} {{Quantum field
  theory and critical phenomena}}}},\ \bibinfo {edition} {4th}\ ed.\ (\bibinfo
  {publisher} {Oxford University Press},\ \bibinfo {address} {Oxford},\
  \bibinfo {year} {2002})\BibitemShut {NoStop}%
\bibitem [{\citenamefont {Coleman}(1976)}]{Coleman:1976uz}%
  \BibitemOpen
  \bibfield  {author} {\bibinfo {author} {\bibfnamefont {S.~R.}\ \bibnamefont
  {Coleman}},\ }\bibfield  {title} {\bibinfo {title} {{More About the Massive
  Schwinger Model}},\ }\href {https://doi.org/10.1016/0003-4916(76)90280-3}
  {\bibfield  {journal} {\bibinfo  {journal} {Ann. Phys. (N.Y.)}\ }\textbf
  {\bibinfo {volume} {101}},\ \bibinfo {pages} {239} (\bibinfo {year}
  {1976})}\BibitemShut {NoStop}%
\bibitem [{\citenamefont {Naon}(1985)}]{Naon:1984zp}%
  \BibitemOpen
  \bibfield  {author} {\bibinfo {author} {\bibfnamefont {C.}~\bibnamefont
  {Naon}},\ }\bibfield  {title} {\bibinfo {title} {{Abelian and Nonabelian
  Bosonization in the Path Integral Framework}},\ }\href
  {https://doi.org/10.1103/PhysRevD.31.2035} {\bibfield  {journal} {\bibinfo
  {journal} {Phys. Rev. D}\ }\textbf {\bibinfo {volume} {31}},\ \bibinfo
  {pages} {2035} (\bibinfo {year} {1985})}\BibitemShut {NoStop}%
\bibitem [{\citenamefont {Coleman}(1975)}]{Coleman:1974bu}%
  \BibitemOpen
  \bibfield  {author} {\bibinfo {author} {\bibfnamefont {S.~R.}\ \bibnamefont
  {Coleman}},\ }\bibfield  {title} {\bibinfo {title} {{The Quantum Sine-Gordon
  Equation as the Massive Thirring Model}},\ }\href
  {https://doi.org/10.1103/PhysRevD.11.2088} {\bibfield  {journal} {\bibinfo
  {journal} {Phys. Rev. D}\ }\textbf {\bibinfo {volume} {11}},\ \bibinfo
  {pages} {2088} (\bibinfo {year} {1975})}\BibitemShut {NoStop}%
\bibitem [{\citenamefont {Faruk}(2018)}]{Faruk:2018mcs}%
  \BibitemOpen
  \bibfield  {author} {\bibinfo {author} {\bibfnamefont {M.~M.}\ \bibnamefont
  {Faruk}},\ }\bibfield  {title} {\bibinfo {title} {{Duality between the
  massive sine-Gordon and the massive Schwinger models at finite
  temperature}},\ }\href {https://doi.org/10.1140/epjp/i2018-12334-1}
  {\bibfield  {journal} {\bibinfo  {journal} {Eur. Phys. J. Plus}\ }\textbf
  {\bibinfo {volume} {133}},\ \bibinfo {pages} {479} (\bibinfo {year}
  {2018})}\BibitemShut {NoStop}%
\bibitem [{\citenamefont {Coleman}\ \emph {et~al.}(1975)\citenamefont
  {Coleman}, \citenamefont {Jackiw},\ and\ \citenamefont
  {Susskind}}]{ColemanJackiw:1975pw}%
  \BibitemOpen
  \bibfield  {author} {\bibinfo {author} {\bibfnamefont {S.~R.}\ \bibnamefont
  {Coleman}}, \bibinfo {author} {\bibfnamefont {R.}~\bibnamefont {Jackiw}},\
  and\ \bibinfo {author} {\bibfnamefont {L.}~\bibnamefont {Susskind}},\
  }\bibfield  {title} {\bibinfo {title} {{Charge Shielding and Quark
  Confinement in the Massive Schwinger Model}},\ }\href
  {https://doi.org/10.1016/0003-4916(75)90212-2} {\bibfield  {journal}
  {\bibinfo  {journal} {Ann. Phys. (N.Y.)}\ }\textbf {\bibinfo {volume} {93}},\
  \bibinfo {pages} {267} (\bibinfo {year} {1975})}\BibitemShut {NoStop}%
\bibitem [{\citenamefont {Frishman}\ and\ \citenamefont
  {Sonnenschein}(2010)}]{frishman_sonnenschein_2010}%
  \BibitemOpen
  \bibfield  {author} {\bibinfo {author} {\bibfnamefont {Y.}~\bibnamefont
  {Frishman}}\ and\ \bibinfo {author} {\bibfnamefont {J.}~\bibnamefont
  {Sonnenschein}},\ }\href@noop {} {\emph {\bibinfo {title} {Non-Perturbative
  Field Theory: From Two Dimensional Conformal Field Theory to QCD in Four
  Dimensions}}},\ Cambridge Monographs on Mathematical Physics\ (\bibinfo
  {publisher} {Cambridge University Press, Cambridge, England},\ \bibinfo
  {year} {2010})\BibitemShut {NoStop}%
\bibitem [{\citenamefont {Hamer}\ \emph {et~al.}(1982)\citenamefont {Hamer},
  \citenamefont {Kogut}, \citenamefont {Crewther},\ and\ \citenamefont
  {Mazzolini}}]{Hamer:1982mx}%
  \BibitemOpen
  \bibfield  {author} {\bibinfo {author} {\bibfnamefont {C.}~\bibnamefont
  {Hamer}}, \bibinfo {author} {\bibfnamefont {J.~B.}\ \bibnamefont {Kogut}},
  \bibinfo {author} {\bibfnamefont {D.}~\bibnamefont {Crewther}},\ and\
  \bibinfo {author} {\bibfnamefont {M.}~\bibnamefont {Mazzolini}},\ }\bibfield
  {title} {\bibinfo {title} {{The Massive Schwinger Model on a Lattice:
  Background Field, Chiral Symmetry and the String Tension}},\ }\href
  {https://doi.org/10.1016/0550-3213(82)90229-2} {\bibfield  {journal}
  {\bibinfo  {journal} {Nucl. Phys.}\ }\textbf {\bibinfo {volume} {B208}},\
  \bibinfo {pages} {413} (\bibinfo {year} {1982})}\BibitemShut {NoStop}%
\bibitem [{\citenamefont {Schiller}\ and\ \citenamefont
  {Ranft}(1983)}]{Schiller:1983sj}%
  \BibitemOpen
  \bibfield  {author} {\bibinfo {author} {\bibfnamefont {A.}~\bibnamefont
  {Schiller}}\ and\ \bibinfo {author} {\bibfnamefont {J.}~\bibnamefont
  {Ranft}},\ }\bibfield  {title} {\bibinfo {title} {{The Massive Schwinger
  Model on the Lattice Studied via a Local Hamiltonian Monte Carlo Method}},\
  }\href {https://doi.org/10.1016/0550-3213(83)90049-4} {\bibfield  {journal}
  {\bibinfo  {journal} {Nucl. Phys.}\ }\textbf {\bibinfo {volume} {B225}},\
  \bibinfo {pages} {204} (\bibinfo {year} {1983})}\BibitemShut {NoStop}%
\bibitem [{\citenamefont {Byrnes}\ \emph {et~al.}(2002)\citenamefont {Byrnes},
  \citenamefont {Sriganesh}, \citenamefont {Bursill},\ and\ \citenamefont
  {Hamer}}]{Byrnes:2002nv}%
  \BibitemOpen
  \bibfield  {author} {\bibinfo {author} {\bibfnamefont {T.}~\bibnamefont
  {Byrnes}}, \bibinfo {author} {\bibfnamefont {P.}~\bibnamefont {Sriganesh}},
  \bibinfo {author} {\bibfnamefont {R.}~\bibnamefont {Bursill}},\ and\ \bibinfo
  {author} {\bibfnamefont {C.}~\bibnamefont {Hamer}},\ }\bibfield  {title}
  {\bibinfo {title} {{Density matrix renormalization group approach to the
  massive Schwinger model}},\ }\href
  {https://doi.org/10.1103/PhysRevD.66.013002} {\bibfield  {journal} {\bibinfo
  {journal} {Phys. Rev. D}\ }\textbf {\bibinfo {volume} {66}},\ \bibinfo
  {pages} {013002} (\bibinfo {year} {2002})}\BibitemShut {NoStop}%
\bibitem [{\citenamefont {Buyens}\ \emph {et~al.}(2017)\citenamefont {Buyens},
  \citenamefont {Montangero}, \citenamefont {Haegeman}, \citenamefont
  {Verstraete},\ and\ \citenamefont {Van~Acoleyen}}]{Buyens:2017crb}%
  \BibitemOpen
  \bibfield  {author} {\bibinfo {author} {\bibfnamefont {B.}~\bibnamefont
  {Buyens}}, \bibinfo {author} {\bibfnamefont {S.}~\bibnamefont {Montangero}},
  \bibinfo {author} {\bibfnamefont {J.}~\bibnamefont {Haegeman}}, \bibinfo
  {author} {\bibfnamefont {F.}~\bibnamefont {Verstraete}},\ and\ \bibinfo
  {author} {\bibfnamefont {K.}~\bibnamefont {Van~Acoleyen}},\ }\bibfield
  {title} {\bibinfo {title} {{Finite-representation approximation of lattice
  gauge theories at the continuum limit with tensor networks}},\ }\href
  {https://doi.org/10.1103/PhysRevD.95.094509} {\bibfield  {journal} {\bibinfo
  {journal} {Phys. Rev. D}\ }\textbf {\bibinfo {volume} {95}},\ \bibinfo
  {pages} {094509} (\bibinfo {year} {2017})}\BibitemShut {NoStop}%
\bibitem [{\citenamefont {Shimizu}\ and\ \citenamefont
  {Kuramashi}(2014)}]{Shimizu:2014fsa}%
  \BibitemOpen
  \bibfield  {author} {\bibinfo {author} {\bibfnamefont {Y.}~\bibnamefont
  {Shimizu}}\ and\ \bibinfo {author} {\bibfnamefont {Y.}~\bibnamefont
  {Kuramashi}},\ }\bibfield  {title} {\bibinfo {title} {{Critical behavior of
  the lattice Schwinger model with a topological term at $\theta=\pi$ using the
  Grassmann tensor renormalization group}},\ }\href
  {https://doi.org/10.1103/PhysRevD.90.074503} {\bibfield  {journal} {\bibinfo
  {journal} {Phys. Rev. D}\ }\textbf {\bibinfo {volume} {90}},\ \bibinfo
  {pages} {074503} (\bibinfo {year} {2014})}\BibitemShut {NoStop}%
\bibitem [{\citenamefont {Saito}\ \emph {et~al.}(2016)\citenamefont {Saito},
  \citenamefont {Banuls}, \citenamefont {Cichy}, \citenamefont {Cirac},\ and\
  \citenamefont {Jansen}}]{Banuls:2015ryj}%
  \BibitemOpen
  \bibfield  {author} {\bibinfo {author} {\bibfnamefont {H.}~\bibnamefont
  {Saito}}, \bibinfo {author} {\bibfnamefont {M.~C.}\ \bibnamefont {Banuls}},
  \bibinfo {author} {\bibfnamefont {K.}~\bibnamefont {Cichy}}, \bibinfo
  {author} {\bibfnamefont {J.~I.}\ \bibnamefont {Cirac}},\ and\ \bibinfo
  {author} {\bibfnamefont {K.}~\bibnamefont {Jansen}},\ }\bibfield  {title}
  {\bibinfo {title} {{Thermal evolution of the 1-flavour Schwinger model with
  using Matrix Product States}},\ }\href {https://doi.org/10.22323/1.251.0283}
  {\bibfield  {journal} {\bibinfo  {journal} {Proc. Sci.}\ }\textbf {\bibinfo
  {volume} {LATTICE2015}},\ \bibinfo {pages} {283} (\bibinfo {year}
  {2016})}\BibitemShut {NoStop}%
\bibitem [{\citenamefont {Bañuls}\ \emph {et~al.}(2016)\citenamefont
  {Bañuls}, \citenamefont {Cichy}, \citenamefont {Jansen},\ and\ \citenamefont
  {Saito}}]{Banuls:2016lkq}%
  \BibitemOpen
  \bibfield  {author} {\bibinfo {author} {\bibfnamefont {M.~C.}\ \bibnamefont
  {Bañuls}}, \bibinfo {author} {\bibfnamefont {K.}~\bibnamefont {Cichy}},
  \bibinfo {author} {\bibfnamefont {K.}~\bibnamefont {Jansen}},\ and\ \bibinfo
  {author} {\bibfnamefont {H.}~\bibnamefont {Saito}},\ }\bibfield  {title}
  {\bibinfo {title} {{Chiral condensate in the Schwinger model with Matrix
  Product Operators}},\ }\href {https://doi.org/10.1103/PhysRevD.93.094512}
  {\bibfield  {journal} {\bibinfo  {journal} {Phys. Rev. D}\ }\textbf {\bibinfo
  {volume} {93}},\ \bibinfo {pages} {094512} (\bibinfo {year}
  {2016})}\BibitemShut {NoStop}%
\bibitem [{\citenamefont {Buyens}\ \emph {et~al.}(2016)\citenamefont {Buyens},
  \citenamefont {Verstraete},\ and\ \citenamefont
  {Van~Acoleyen}}]{Buyens:2016ecr}%
  \BibitemOpen
  \bibfield  {author} {\bibinfo {author} {\bibfnamefont {B.}~\bibnamefont
  {Buyens}}, \bibinfo {author} {\bibfnamefont {F.}~\bibnamefont {Verstraete}},\
  and\ \bibinfo {author} {\bibfnamefont {K.}~\bibnamefont {Van~Acoleyen}},\
  }\bibfield  {title} {\bibinfo {title} {{Hamiltonian simulation of the
  Schwinger model at finite temperature}},\ }\href
  {https://doi.org/10.1103/PhysRevD.94.085018} {\bibfield  {journal} {\bibinfo
  {journal} {Phys. Rev. D}\ }\textbf {\bibinfo {volume} {94}},\ \bibinfo
  {pages} {085018} (\bibinfo {year} {2016})}\BibitemShut {NoStop}%
\bibitem [{\citenamefont {Zamolodchikov}(1995)}]{Zamolodchikov:1995xk}%
  \BibitemOpen
  \bibfield  {author} {\bibinfo {author} {\bibfnamefont {A.~B.}\ \bibnamefont
  {Zamolodchikov}},\ }\bibfield  {title} {\bibinfo {title} {{Mass scale in the
  sine-Gordon model and its reductions}},\ }\href
  {https://doi.org/10.1142/S0217751X9500053X} {\bibfield  {journal} {\bibinfo
  {journal} {Int. J. Mod. Phys. A}\ }\textbf {\bibinfo {volume} {10}},\
  \bibinfo {pages} {1125} (\bibinfo {year} {1995})}\BibitemShut {NoStop}%
\bibitem [{\citenamefont {Jayewardena}(1988)}]{Jayewardena:1988td}%
  \BibitemOpen
  \bibfield  {author} {\bibinfo {author} {\bibfnamefont {C.}~\bibnamefont
  {Jayewardena}},\ }\bibfield  {title} {\bibinfo {title} {{Schwinger Model on
  $S^2$}},\ }\href@noop {} {\bibfield  {journal} {\bibinfo  {journal} {Helv.
  Phys. Acta}\ }\textbf {\bibinfo {volume} {61}},\ \bibinfo {pages} {636}
  (\bibinfo {year} {1988})}\BibitemShut {NoStop}%
\bibitem [{\citenamefont {Sachs}\ and\ \citenamefont
  {Wipf}(1992)}]{Sachs:1991en}%
  \BibitemOpen
  \bibfield  {author} {\bibinfo {author} {\bibfnamefont {I.}~\bibnamefont
  {Sachs}}\ and\ \bibinfo {author} {\bibfnamefont {A.}~\bibnamefont {Wipf}},\
  }\bibfield  {title} {\bibinfo {title} {{Finite temperature Schwinger
  model}},\ }\href@noop {} {\bibfield  {journal} {\bibinfo  {journal} {Helv.
  Phys. Acta}\ }\textbf {\bibinfo {volume} {65}},\ \bibinfo {pages} {652}
  (\bibinfo {year} {1992})}\BibitemShut {NoStop}%
\bibitem [{not()}]{note1}%
  \BibitemOpen
  \bibinfo {note} {This result can also be understood within the bosonic model
  from the following scaling argument. In the weak-coupling limit
  $u/\Lambda^2,M/\Lambda\ll 1$, the UV part of the flow is controlled by the
  Gaussian fixed point $u=M=0$. Using the fact that $M$ and $u$ have scaling
  dimensions $[M]=[u]=1$ (the latter result follows from $[e^{i\beta\phi}]=1$),
  a quantity $X$ with scaling dimension $[X]=a$ transforms as
  $X(M/\Lambda,u/\Lambda^2) = s^{-a} X(s M/\Lambda,s u/\Lambda^2)$ in a scale
  transformation. Setting $s=\Lambda/M$, this gives $X(M/\Lambda,u/\Lambda^2) =
  (M/\Lambda)^a X(1,u/M\Lambda) \equiv M^a f(u/M\Lambda)$, where
  $f(u/M\Lambda)$ is a universal scaling function.}\BibitemShut {Stop}%
\bibitem [{\citenamefont {Canet}\ \emph {et~al.}(2003)\citenamefont {Canet},
  \citenamefont {Delamotte}, \citenamefont {Mouhanna},\ and\ \citenamefont
  {Vidal}}]{Canet:2002gs}%
  \BibitemOpen
  \bibfield  {author} {\bibinfo {author} {\bibfnamefont {L.}~\bibnamefont
  {Canet}}, \bibinfo {author} {\bibfnamefont {B.}~\bibnamefont {Delamotte}},
  \bibinfo {author} {\bibfnamefont {D.}~\bibnamefont {Mouhanna}},\ and\
  \bibinfo {author} {\bibfnamefont {J.}~\bibnamefont {Vidal}},\ }\bibfield
  {title} {\bibinfo {title} {{Optimization of the derivative expansion in the
  nonperturbative renormalization group}},\ }\href
  {https://doi.org/10.1103/PhysRevD.67.065004} {\bibfield  {journal} {\bibinfo
  {journal} {Phys. Rev. D}\ }\textbf {\bibinfo {volume} {67}},\ \bibinfo
  {pages} {065004} (\bibinfo {year} {2003})}\BibitemShut {NoStop}%
\bibitem [{\citenamefont {Pelissetto}\ and\ \citenamefont
  {Vicari}(2002)}]{Pelissetto:2000ek}%
  \BibitemOpen
  \bibfield  {author} {\bibinfo {author} {\bibfnamefont {A.}~\bibnamefont
  {Pelissetto}}\ and\ \bibinfo {author} {\bibfnamefont {E.}~\bibnamefont
  {Vicari}},\ }\bibfield  {title} {\bibinfo {title} {{Critical phenomena and
  renormalization group theory}},\ }\href
  {https://doi.org/10.1016/S0370-1573(02)00219-3} {\bibfield  {journal}
  {\bibinfo  {journal} {Phys. Rep.}\ }\textbf {\bibinfo {volume} {368}},\
  \bibinfo {pages} {549} (\bibinfo {year} {2002})}\BibitemShut {NoStop}%
\bibitem [{\citenamefont {Tetradis}\ and\ \citenamefont
  {Wetterich}(1992)}]{Tetradis92}%
  \BibitemOpen
  \bibfield  {author} {\bibinfo {author} {\bibfnamefont {N.}~\bibnamefont
  {Tetradis}}\ and\ \bibinfo {author} {\bibfnamefont {C.}~\bibnamefont
  {Wetterich}},\ }\bibfield  {title} {\bibinfo {title} {{Scale dependence of
  the average potential around the maximum in $\phi^4$ theories}},\ }\href
  {https://doi.org/doi:10.1016/0550-3213(92)90676-3} {\bibfield  {journal}
  {\bibinfo  {journal} {Nucl. Phys.}\ }\textbf {\bibinfo {volume} {B383}},\
  \bibinfo {pages} {197} (\bibinfo {year} {1992})}\BibitemShut {NoStop}%
\bibitem [{\citenamefont {Tetradis}\ and\ \citenamefont
  {Litim}(1996)}]{Tetradis96}%
  \BibitemOpen
  \bibfield  {author} {\bibinfo {author} {\bibfnamefont {N.}~\bibnamefont
  {Tetradis}}\ and\ \bibinfo {author} {\bibfnamefont {D.~F.}\ \bibnamefont
  {Litim}},\ }\bibfield  {title} {\bibinfo {title} {{Analytical solutions of
  exact renormalization group equations }},\ }\href
  {https://doi.org/http://dx.doi.org/10.1016/0550-3213(95)00642-7} {\bibfield
  {journal} {\bibinfo  {journal} {Nucl. Phys.}\ }\textbf {\bibinfo {volume}
  {B464}},\ \bibinfo {pages} {492} (\bibinfo {year} {1996})}\BibitemShut
  {NoStop}%
\bibitem [{\citenamefont {Pel\'aez}\ and\ \citenamefont
  {Wschebor}(2016)}]{Pelaez:2015nsa}%
  \BibitemOpen
  \bibfield  {author} {\bibinfo {author} {\bibfnamefont {M.}~\bibnamefont
  {Pel\'aez}}\ and\ \bibinfo {author} {\bibfnamefont {N.}~\bibnamefont
  {Wschebor}},\ }\bibfield  {title} {\bibinfo {title} {{Ordered phase of the
  $O(N)$ model within the nonperturbative renormalization group}},\ }\href
  {https://doi.org/10.1103/PhysRevE.94.042136} {\bibfield  {journal} {\bibinfo
  {journal} {Phys. Rev. E}\ }\textbf {\bibinfo {volume} {94}},\ \bibinfo
  {pages} {042136} (\bibinfo {year} {2016})}\BibitemShut {NoStop}%
\bibitem [{\citenamefont {Debelhoir}\ and\ \citenamefont
  {Dupuis}(2016)}]{Debelhoir16b}%
  \BibitemOpen
  \bibfield  {author} {\bibinfo {author} {\bibfnamefont {T.}~\bibnamefont
  {Debelhoir}}\ and\ \bibinfo {author} {\bibfnamefont {N.}~\bibnamefont
  {Dupuis}},\ }\bibfield  {title} {\bibinfo {title} {{First-order phase
  transitions in spinor Bose gases and frustrated magnets}},\ }\href
  {https://doi.org/10.1103/PhysRevA.94.053623} {\bibfield  {journal} {\bibinfo
  {journal} {Phys. Rev. A}\ }\textbf {\bibinfo {volume} {94}},\ \bibinfo
  {pages} {053623} (\bibinfo {year} {2016})}\BibitemShut {NoStop}%
\bibitem [{\citenamefont {Funcke}\ \emph {et~al.}(2020)\citenamefont {Funcke},
  \citenamefont {Jansen},\ and\ \citenamefont {K\"uhn}}]{funcke}%
  \BibitemOpen
  \bibfield  {author} {\bibinfo {author} {\bibfnamefont {L.}~\bibnamefont
  {Funcke}}, \bibinfo {author} {\bibfnamefont {K.}~\bibnamefont {Jansen}},\
  and\ \bibinfo {author} {\bibfnamefont {S.}~\bibnamefont {K\"uhn}},\
  }\bibfield  {title} {\bibinfo {title} {Topological vacuum structure of the
  schwinger model with matrix product states},\ }\href
  {https://doi.org/10.1103/PhysRevD.101.054507} {\bibfield  {journal} {\bibinfo
   {journal} {Phys. Rev. D}\ }\textbf {\bibinfo {volume} {101}},\ \bibinfo
  {pages} {054507} (\bibinfo {year} {2020})}\BibitemShut {NoStop}%
\bibitem [{\citenamefont {Nandori}\ \emph {et~al.}(2014)\citenamefont
  {Nandori}, \citenamefont {Marian},\ and\ \citenamefont
  {Bacso}}]{Nandori:2013nda}%
  \BibitemOpen
  \bibfield  {author} {\bibinfo {author} {\bibfnamefont {I.}~\bibnamefont
  {Nandori}}, \bibinfo {author} {\bibfnamefont {I.~G.}\ \bibnamefont
  {Marian}},\ and\ \bibinfo {author} {\bibfnamefont {V.}~\bibnamefont
  {Bacso}},\ }\bibfield  {title} {\bibinfo {title} {{Spontaneous symmetry
  breaking and optimization of functional renormalization group}},\ }\href
  {https://doi.org/10.1103/PhysRevD.89.047701} {\bibfield  {journal} {\bibinfo
  {journal} {Phys. Rev. D}\ }\textbf {\bibinfo {volume} {89}},\ \bibinfo
  {pages} {047701} (\bibinfo {year} {2014})}\BibitemShut {NoStop}%
\bibitem [{\citenamefont {Rajaraman}(1989)}]{Rajaraman_book}%
  \BibitemOpen
  \bibfield  {author} {\bibinfo {author} {\bibfnamefont {R.}~\bibnamefont
  {Rajaraman}},\ }\href@noop {} {\emph {\bibinfo {title} {{Solitons and
  instantons}}}}\ (\bibinfo  {publisher} {North-Holland},\ \bibinfo {address}
  {Amsterdam},\ \bibinfo {year} {1989})\BibitemShut {NoStop}%
\bibitem [{\citenamefont {Komargodski}\ \emph {et~al.}(2021)\citenamefont
  {Komargodski}, \citenamefont {Ohmori}, \citenamefont {Roumpedakis},\ and\
  \citenamefont {Seifnashri}}]{Komargodski2020SymmetriesQCD2}%
  \BibitemOpen
  \bibfield  {author} {\bibinfo {author} {\bibfnamefont {Z.}~\bibnamefont
  {Komargodski}}, \bibinfo {author} {\bibfnamefont {K.}~\bibnamefont {Ohmori}},
  \bibinfo {author} {\bibfnamefont {K.}~\bibnamefont {Roumpedakis}},\ and\
  \bibinfo {author} {\bibfnamefont {S.}~\bibnamefont {Seifnashri}},\ }\bibfield
   {title} {\bibinfo {title} {{Symmetries and strings of adjoint QCD2}},\
  }\href {https://doi.org/10.1007/JHEP03(2021)103} {\bibfield  {journal}
  {\bibinfo  {journal} {{J. High Energy Phys.}}\ }\textbf {\bibinfo {volume}
  {03}},\ \bibinfo {pages} {103} (\bibinfo {year} {2021})}\BibitemShut
  {NoStop}%
\bibitem [{\citenamefont {Komargodski}\ \emph {et~al.}(2019)\citenamefont
  {Komargodski}, \citenamefont {Sharon}, \citenamefont {Thorngren},\ and\
  \citenamefont {Zhou}}]{Komargodski2}%
  \BibitemOpen
  \bibfield  {author} {\bibinfo {author} {\bibfnamefont {Z.}~\bibnamefont
  {Komargodski}}, \bibinfo {author} {\bibfnamefont {A.}~\bibnamefont {Sharon}},
  \bibinfo {author} {\bibfnamefont {R.}~\bibnamefont {Thorngren}},\ and\
  \bibinfo {author} {\bibfnamefont {X.}~\bibnamefont {Zhou}},\ }\bibfield
  {title} {\bibinfo {title} {{Comments on Abelian Higgs Models and Persistent
  Order}},\ }\href {https://doi.org/10.21468/SciPostPhys.6.1.003} {\bibfield
  {journal} {\bibinfo  {journal} {SciPost Phys.}\ }\textbf {\bibinfo {volume}
  {6}},\ \bibinfo {pages} {3} (\bibinfo {year} {2019})}\BibitemShut {NoStop}%
\bibitem [{\citenamefont {Dupuis}\ and\ \citenamefont
  {Daviet}(2020)}]{Dupuis20}%
  \BibitemOpen
  \bibfield  {author} {\bibinfo {author} {\bibfnamefont {N.}~\bibnamefont
  {Dupuis}}\ and\ \bibinfo {author} {\bibfnamefont {R.}~\bibnamefont
  {Daviet}},\ }\bibfield  {title} {\bibinfo {title} {Bose-glass phase of a
  one-dimensional disordered bose fluid: Metastable states, quantum tunneling,
  and droplets},\ }\href {https://doi.org/10.1103/PhysRevE.101.042139}
  {\bibfield  {journal} {\bibinfo  {journal} {Phys. Rev. E}\ }\textbf {\bibinfo
  {volume} {101}},\ \bibinfo {pages} {042139} (\bibinfo {year}
  {2020})}\BibitemShut {NoStop}%
\bibitem [{\citenamefont {Daviet}\ and\ \citenamefont
  {Dupuis}(2020)}]{Daviet20}%
  \BibitemOpen
  \bibfield  {author} {\bibinfo {author} {\bibfnamefont {R.}~\bibnamefont
  {Daviet}}\ and\ \bibinfo {author} {\bibfnamefont {N.}~\bibnamefont
  {Dupuis}},\ }\bibfield  {title} {\bibinfo {title} {Mott-glass phase of a
  one-dimensional quantum fluid with long-range interactions},\ }\href
  {https://doi.org/10.1103/PhysRevLett.125.235301} {\bibfield  {journal}
  {\bibinfo  {journal} {Phys. Rev. Lett.}\ }\textbf {\bibinfo {volume} {125}},\
  \bibinfo {pages} {235301} (\bibinfo {year} {2020})}\BibitemShut {NoStop}%
\bibitem [{\citenamefont {Daviet}\ and\ \citenamefont
  {Dupuis}(2021)}]{Daviet21}%
  \BibitemOpen
  \bibfield  {author} {\bibinfo {author} {\bibfnamefont {R.}~\bibnamefont
  {Daviet}}\ and\ \bibinfo {author} {\bibfnamefont {N.}~\bibnamefont
  {Dupuis}},\ }\bibfield  {title} {\bibinfo {title} {Chaos in the bose-glass
  phase of a one-dimensional disordered bose fluid},\ }\href
  {https://doi.org/10.1103/PhysRevE.103.052136} {\bibfield  {journal} {\bibinfo
   {journal} {Phys. Rev. E}\ }\textbf {\bibinfo {volume} {103}},\ \bibinfo
  {pages} {052136} (\bibinfo {year} {2021})}\BibitemShut {NoStop}%
\bibitem [{\citenamefont {Motrunich}\ and\ \citenamefont
  {Vishwanath}(2004)}]{Motrunich04}%
  \BibitemOpen
  \bibfield  {author} {\bibinfo {author} {\bibfnamefont {O.~I.}\ \bibnamefont
  {Motrunich}}\ and\ \bibinfo {author} {\bibfnamefont {A.}~\bibnamefont
  {Vishwanath}},\ }\bibfield  {title} {\bibinfo {title} {Emergent photons and
  transitions in the $\mathrm{O}(3)$ sigma model with hedgehog suppression},\
  }\href {https://doi.org/10.1103/PhysRevB.70.075104} {\bibfield  {journal}
  {\bibinfo  {journal} {Phys. Rev. B}\ }\textbf {\bibinfo {volume} {70}},\
  \bibinfo {pages} {075104} (\bibinfo {year} {2004})}\BibitemShut {NoStop}%
\bibitem [{\citenamefont {Chlebicki}\ and\ \citenamefont
  {Jakubczyk}(2021)}]{Chlebicki21}%
  \BibitemOpen
  \bibfield  {author} {\bibinfo {author} {\bibfnamefont {A.}~\bibnamefont
  {Chlebicki}}\ and\ \bibinfo {author} {\bibfnamefont {P.}~\bibnamefont
  {Jakubczyk}},\ }\bibfield  {title} {\bibinfo {title} {{Analyticity of
  critical exponents of the $O(N)$ models from nonperturbative
  renormalization}},\ }\href {https://doi.org/10.21468/SciPostPhys.10.6.134}
  {\bibfield  {journal} {\bibinfo  {journal} {SciPost Phys.}\ }\textbf
  {\bibinfo {volume} {10}},\ \bibinfo {pages} {134} (\bibinfo {year}
  {2021})}\BibitemShut {NoStop}%
\bibitem [{\citenamefont {Rulquin}\ \emph {et~al.}(2016)\citenamefont
  {Rulquin}, \citenamefont {Urbani}, \citenamefont {Biroli}, \citenamefont
  {Tarjus},\ and\ \citenamefont {Tarzia}}]{Rulquin15a}%
  \BibitemOpen
  \bibfield  {author} {\bibinfo {author} {\bibfnamefont {C.}~\bibnamefont
  {Rulquin}}, \bibinfo {author} {\bibfnamefont {P.}~\bibnamefont {Urbani}},
  \bibinfo {author} {\bibfnamefont {G.}~\bibnamefont {Biroli}}, \bibinfo
  {author} {\bibfnamefont {G.}~\bibnamefont {Tarjus}},\ and\ \bibinfo {author}
  {\bibfnamefont {M.}~\bibnamefont {Tarzia}},\ }\bibfield  {title} {\bibinfo
  {title} {Nonperturbative fluctuations and metastability in a simple model:
  from observables to microscopic theory and back},\ }\href
  {https://doi.org/10.1088/1742-5468/2016/02/023209} {\bibfield  {journal}
  {\bibinfo  {journal} {Journal of Statistical Mechanics: Theory and
  Experiment}\ }\textbf {\bibinfo {volume} {2016}},\ \bibinfo {pages} {023209}
  (\bibinfo {year} {2016})}\BibitemShut {NoStop}%
\bibitem [{\citenamefont {Jakubczyk}\ \emph {et~al.}(2014)\citenamefont
  {Jakubczyk}, \citenamefont {Dupuis},\ and\ \citenamefont
  {Delamotte}}]{Jakubczyk14}%
  \BibitemOpen
  \bibfield  {author} {\bibinfo {author} {\bibfnamefont {P.}~\bibnamefont
  {Jakubczyk}}, \bibinfo {author} {\bibfnamefont {N.}~\bibnamefont {Dupuis}},\
  and\ \bibinfo {author} {\bibfnamefont {B.}~\bibnamefont {Delamotte}},\
  }\bibfield  {title} {\bibinfo {title} {{Reexamination of the nonperturbative
  renormalization-group approach to the Kosterlitz-Thouless transition}},\
  }\href {https://doi.org/10.1103/PhysRevE.90.062105} {\bibfield  {journal}
  {\bibinfo  {journal} {Phys. Rev. E}\ }\textbf {\bibinfo {volume} {90}},\
  \bibinfo {pages} {062105} (\bibinfo {year} {2014})}\BibitemShut {NoStop}%
\end{thebibliography}%


\end{document}